\documentstyle[psfig,emulateapj]{article}

\lefthead{Barger et al.}

\begin{document}
\title{Mapping the evolution of high redshift dusty galaxies with
submillimeter observations of a radio-selected sample
}
\author{A.\,J.\ Barger,$\!$\altaffilmark{1,2,3,4}
L.\,L.\ Cowie,$\!$\altaffilmark{1,4}
E.\,A.\ Richards,\altaffilmark{2,4,5}
}

\altaffiltext{1}{Institute for Astronomy, University of Hawaii,
2680 Woodlawn Drive, Honolulu, Hawaii 96822}
\altaffiltext{2}{Hubble Fellow}
\altaffiltext{3}{Chandra Fellow at Large}
\altaffiltext{4}{Visiting Astronomer, W.\,M.\ Keck Observatory, jointly
operated by the California Institute of Technology and the University
of California}
\altaffiltext{5}{National Radio Astronomy Observatory \& University of
Virginia, 520 Edgemont Road, Charlottesville, VA 22903}

\slugcomment{Accepted by the Astronomical Journal for the April 2000 issue}

\begin{abstract}
Direct submillimeter 
imaging has recently revealed the 850\ $\mu$m background to be mostly composed 
of a population of distant ultraluminous infrared galaxies, but
identifying the optical/near-infrared (NIR) counterparts
to these sources has proved difficult due to the poor submillimeter 
spatial resolution.
However, the proportionality of both centimeter and submillimeter
data to the star formation rate suggests that high resolution radio
continuum maps with subarcsecond positional accuracy can be 
exploited to locate submillimeter sources.
In this paper we present results from a targeted SCUBA survey of 
micro-Jansky radio sources in the flanking fields of the Hubble 
Deep Field. The sources were selected from the uniform 
(8\ $\mu$Jy at 1$\sigma$) 1.4\ GHz VLA image of Richards (1999b).
Even with relatively shallow SCUBA observations (a $3\sigma$ detection
limit of 6\ mJy at 850\ $\mu$m),
we were successful at making submillimeter detections of
optical/NIR-faint ($I\gtrsim 24$ and $K\simeq 21-22$) 
radio sources, and our counts closely 
match the bright counts from submillimeter surveys.
An important corollary is that a large fraction of the bright 
($>6$\ mJy) submillimeter sources in untargeted 
submillimeter surveys have extremely faint optical/NIR
counterparts and hence are inaccessible to optical imaging and spectroscopy.
However, redshift estimates can be made from the ratio of the
submillimeter flux to the radio flux across the 100 GHz break in the
spectral energy distribution.
This procedure, which we refer to as {\em millimetric}
redshift estimation, places the bright submillimeter population
at $z=1-3$ where it forms the high redshift tail of the faint radio
population. The star formation rate density (SFRD) due to ultraluminous
infrared galaxies increases by more than two orders of magnitude 
from $z\sim 0$ to $z\sim 1-3$. The SFRD at high redshift inferred from 
our $>6$\ mJy submillimeter observations is comparable to that observed
in the ultraviolet/optical.
\end{abstract}


\section{Introduction}
\label{secintro}

Recent detections of distant dusty galaxies with the SCUBA camera 
(the Submillimeter Common User Bolometer Array;
\markcite{holland98}Holland et al.\ 1999)
on the 15\ m James Clerk Maxwell Telescope\footnote{The JCMT is operated
by the Joint Astronomy Center on behalf of the parent organizations,
the Particle Physics and Astronomy Research Council in the United
Kingdom, the National Research Council of Canada, and the Netherlands
Organization for Scientific Research.}
constitute a substantial fraction of
the cosmic FIR background detected
by the {\it FIRAS} and {\it DIRBE} experiments on the 
{\it COBE} satellite (\markcite{puget96}Puget et al.\ 1996;
\markcite{guider97}Guiderdoni et al.\ 1997;
\markcite{schlegel97}Schlegel, Finkbeiner, \& Davis\ 1998; 
\markcite{fixsen98}Fixsen
et al.\ 1998; \markcite{hauser98}Hauser et al.\ 1998;
\markcite{lagache99}Lagache et al.\ 1999). Since the
observed FIR background is comparable to the total 
unobscured emission at ultraviolet/optical wavelengths, 
a full determination of the global star formation history of the Universe 
requires a comprehensive understanding of this
dust-enshrouded galaxy population.
The 850\ $\mu$m SCUBA surveys to date have reported galaxy number counts
that are in general agreement
(\markcite{smail97}Smail, Ivison \& Blain 1997;
\markcite{barger98}Barger et al.\ 1998;
\markcite{hughes98}Hughes et al.\ 1998;
\markcite{blain99a}Blain et al.\ 1999a;
\markcite{eales99}Eales et al.\ 1999;
\markcite{barger99}Barger, Cowie, \& Sanders 1999);
the cumulative surface density above 2\ mJy is about 
$3\times 10^3\ {\rm deg}^{-2}$.
The discrete sources have bolometric luminosities that are
characteristically $\gtrsim 10^{12}\ h_{65}^{-2}\ {\rm L}_\odot$
if they lie at $z\gtrsim 1$. Moreover, the sources for which
measurements exist at multiple wavelengths (e.g.\ Ivison et al.\ 1998)
show thermal spectral energy distributions (SEDs). Thus,
the SCUBA sources are inferred to be the distant analogs of the
local ultraluminous infrared galaxy (ULIG;
\markcite{sanders96}Sanders \& Mirabel 1996) population.

An essential observational goal is to determine the redshift distribution 
of the submillimeter population in order to trace the extent and evolution 
of obscured emission in the distant Universe. However,
identifying the optical/NIR counterparts to the submillimeter sources is 
difficult due to the uncertainty in the SCUBA positions.
\markcite{barger99b}Barger et al.\ (1999b) 
presented a spectroscopic survey of possible optical counterparts 
to a flux-limited sample of galaxies selected from the 850\ $\mu$m 
survey of massive lensing clusters by 
\markcite{smail98}Smail et al.\ (1998). 
Candidate optical counterparts in the SCUBA error-boxes were identified
using moderately deep ground-based and {\it HST} exposures
($I\sim 23.5$ and $I\sim 26$, respectively). One-quarter of the sources
could be reliably identified, and those had redshifts in the range
$z\sim 1-3$. A lower limit of 20\ per cent of the full sample showed
signs of AGN activity. However, for the majority of the submillimeter
sources there were either no optical counterparts or the optical
associations were not secure. Such sources could either be at very high 
redshift or be so highly obscured that they emit their energy
almost entirely in the submillimeter.

High resolution radio continuum maps with subarcsecond positional 
accuracy and resolution offer new opportunities for locating 
submillimeter sources and determining their physical properties.
The unique advantage of centimeter and FIR observations is that
galaxies and the intergalactic medium are transparent at these
wavelengths, so observed flux densities are
proportional to intrinsic luminosities.
In galaxies without a powerful AGN, the radio luminosity is dominated 
by diffuse synchrotron emission from relativistic electrons accelerated
in supernovae remnants from stars more massive than $8\ {\rm M}_\odot$.
These massive stars live $\lesssim 3\times 10^7$\ yr; the relativistic
electrons probably live $\lesssim 10^8$\ yr (\markcite{condon92}Condon 1992).
Thus, radio observations probe very recent star formation.
FIR observations of starburst galaxies are also a direct
measure of massive star formation.
As summarized by \markcite{condon92}Condon (1992), 
radio continuum emission and thermal 
dust emission are empirically observed to be tightly correlated
due to both being linearly related to the massive star formation rate.
If the FIR-radio correlation applies to high
redshift, as is plausibly the case (though at the very highest
redshifts Compton cooling of the relativistic electrons by
the microwave background may suppress the radio emission), then very sensitive
radio observations can be used to pinpoint distant submillimeter sources.

In this paper we investigate the feasibility of using radio data 
to identify and characterize the bright submillimeter source population.
\markcite{richards99b}Richards (1999b) 
recently obtained an extremely deep Very Large Array (VLA) 
1.4\ GHz image centered on the Hubble Deep Field (HDF).
\markcite{r99}Richards et al.\ (1999) matched
ground-based optical data from
\markcite{barger99a}Barger et al.\ (1999a)
to the radio image and found that $\sim 20$ per cent of the galaxies 
in the sample could not be identified to
optical magnitude limits of $I\sim 25$.
In other respects, such as radio size and spectral index, 
the optically-faint objects were not any 
different from the remaining population. Richards et al.\ (1999)
proposed four possible scenarios to explain this
population, including $1<z<3$ obscured starbursts
(beyond $z\sim 3$ the sensitivity to star forming
galaxies cuts off due to the flux density limits of the radio data),
extreme redshift ($z>6$) AGN, $z<2$ obscured AGN,
or one-sided radio jets.

In the first phase of our program we observed with LRIS on
the Keck~II 10\ m telescope a complete subsample of the radio sources.
Our primary objective was to determine the redshifts
of the optical/NIR-faint ($HK'>20.5$) radio sources. Although
we were able to spectroscopically identify nearly all the objects in
our subsample to $HK'\lesssim 20$ (all had $z\lesssim 1.3$),
we were unable to obtain redshifts for the fainter objects.
Either these sources are distant ($z>1.5$) with spectral
features lying outside the optical wavelength range, 
or the visibility of remarkable
features is strongly affected by dust in the galaxies.

If the optical/NIR-faint radio sources are highly dust obscured
systems, then it is possible that they will be detectable in the 
submillimeter. In the second phase of our program we observed with
SCUBA 15 of the 22 optical/NIR-faint radio sources in the central
$\sim 80$\ square arcminute region of the radio map; another 4 were 
observed by Hughes et al.\ (1998; hereafter H98) in the HDF-proper. 
The jiggle map mode enabled simultaneous observations of a
large fraction (31/48) of the optical/NIR-bright radio sources.
Even with relatively shallow SCUBA observations (a $3\sigma$ detection
limit of 6\ mJy at 850\ $\mu$m), we detected 5 of the
optical/NIR-faint radio sources; a sixth
with submillimeter flux $<6$\ mJy was detected in the deep HDF-proper
submillimeter map of H98. 
In contrast, none of the optical/NIR-bright radio sources were detected. 
We additionally detected two $>6$\ mJy sources
that did not have radio counterparts. Thus,
our targeted SCUBA survey of optical/NIR-faint 
radio sources turned up $\sim 70$ per cent 
of the bright submillimeter sources in our surveyed areas.

In the final phase of our program, we explored the feasibility of
obtaining redshift estimates from the submillimeter-to-radio flux ratios,
as recently suggested by \markcite{cy99}Carilli \& Yun (1999).
We find that the redshifted Arp~220 SED
reasonably describes both redshifted local ULIG data
and known high redshift submillimeter sources and hence can
be used as a rough redshift estimator. We estimate that all 
of our bright submillimeter sources fall in
the redshift range $z=1-3$, consistent with the redshifts for the
lensed submillimeter sources of
\markcite{barger99b}Barger et al.\ (1999b).

Once we have redshifts for the 
distant submillimeter source population, we can determine the
global evolution of star formation in dust-obscured galaxies.
Previous studies of the star formation rate density 
(SFRD) have primarily used
rest-frame ultraviolet data (e.g.\ \markcite{madau96}Madau et al.\ 1996).
However, the ultraviolet emission from a galaxy is heavily affected 
by the presence of even small amounts of dust, and the extinction 
corrections are highly uncertain; for example, corrections for
dust obscuration at $z\approx 3$ range from factors of $\sim 3$
(\markcite{pettini97}Pettini et al.\ 1997) 
to factors of $\sim 15$ (\markcite{meurer97}Meurer et al.\ 1997),
though the more recent estimate of 
\markcite{meurer99}Meurer, Heckman, \& Calzetti (1999) 
lowers the latter to a factor of $\sim 5$.
Only direct measurements of the reradiated light at submillimeter 
wavelengths can securely address the SFRD at high redshifts.

The true SFRD will receive contributions from both 
the ultraviolet/optical and the submillimeter. We find that
the submillimeter contribution to the SFRD in the $z=1-3$ range 
from our $>6$\ mJy sources is comparable to the
observed ultraviolet/optical SFRD contribution.
The ultraviolet Lyman-break galaxies are on the average undetected
in the submillimeter at a $1\sigma$ level of $\sim 0.5$\ mJy, and
thus the Lyman-break galaxy population is largely distinct from
the bright submillimeter population (\markcite{chapman99}Chapman et al.\ 1999).
If we assume that fainter submillimeter sources have the same
redshift distribution and properties as the $>6$\ mJy sample,
then the SFRD of the entire population contributing to the
submillimeter background is about an order of magnitude higher
than the observed ultraviolet/optical SFRD.
The contribution from ULIGs to the SFRD increases by more than
two orders of magnitude from $z\sim 0$ to $z=1-3$, which supports 
a scenario in which the distant submillimeter sources are the 
progenitors of massive spheroidal systems. 
This rapid evolution in the ULIG
population sampled by SCUBA can be reproduced in models
(e.g., \markcite{blain99b}Blain et al.\ 1999b,c). 

In \S~2 we present our radio sample and optical/NIR imaging, along 
with our new SCUBA and LRIS observations. In
\S~3 we determine our radio-selected submillimeter source counts.
In \S~4 we introduce the predicted high redshift 
submillimeter-radio flux correlation and obtain millimetric
redshift estimates for our sources. In \S~5
we use the complementary information from radio and submillimeter 
fluxes to gain insights into the characteristics of
our radio-selected submillimeter source population. 
In \S~6 we compare the rest-frame colors of the radio sources
that have spectroscopic redshifts with the rest-frame colors of our
submillimeter sources and find that the latter
are likely to fall in the extremely red object category.
In \S~7 we calculate the
luminosities, number densities, and SFRDs
of our submillimeter sources. We compare the submillimeter 
contribution to the SFRD with contributions
from radio and ultraviolet/optical wavebands over a range of redshifts.
In \S~8 we summarize our main conclusions.
We take ${\rm H_o}=65\ h_{65}$\ km\ s$^{-1}$\ Mpc$^{-1}$ and 
consider both $\Omega_{\rm M}=1$, $\Omega_\Lambda=0$ and
$\Omega_{\rm M}=1/3$, $\Omega_\Lambda=2/3$, which should cover
the full range of possible cosmologies.

\section{Samples and Observations}
\label{secdata}

The present study is based on deep radio maps centered on the 
HDF that were observed with the VLA at 
1.4\ GHz (Richards 1999b)
and 8.5\ GHz (\markcite{richards98}Richards et al.\ 1998). The primary 
1.4\ GHz image covers a $40'$ diameter region with an effective
resolution of $1.8''$ and a $5\sigma$ completeness limit of 40\ $\mu$Jy.
The 8.5\ GHz images have an effective resolution of $3.5''$ and a
$5\sigma$ completeness limit of 8\ $\mu$Jy over a radius of
$\sim 1'$ from the HDF center, rising to 40\ $\mu$Jy at $6.6'$. 

The 1.4\ GHz HDF map was trimmed to match the 79.4\ arcmin$^2$
NIR and optical imaging of the field obtained by
\markcite{barger99a}Barger et al.\ (1999a). 
The absolute radio positions are known to $0.1''-0.2''$ rms
in the HDF; the alignment of the optical data to the radio data
left residual astrometric uncertainties of $\sim0.2''$
(Richards et al.\ 1999).
There are 70 sources in our final radio sample.

\begin{table}
\dummytable\label{tab1}
\end{table}

Table~1 gives the 1.4\ GHz radio catalog for the central
79.4\ arcmin$^2$ region discussed above. 
The first five columns are 
catalog number, RA(2000), Dec(2000),
1.4\ GHz radio flux, and radio flux uncertainty 
(Richards 1999b).
The remaining columns, to be discussed in the following
subsections, are $HK'$, $I$, $V$, $R$, $B$, and $U'$\ magnitudes, 
redshift, 
submillimeter flux, submillimeter flux uncertainty, and
$6.75$ and 15\ $\mu$m ISOCAM fluxes
(\markcite{aussel}Aussel et al.\ 1999). 
The last column of Table~1 gives the radio spectral index $\alpha_r$
($S\propto \nu^{\alpha_r}$) or limit for the sources where 
8.4\ GHz data are available and can be used
to determine the origin of the radio emission (Richards 1999b).
Inverted spectrum sources ($\alpha_r>0$) invariably have self-absorbed 
synchrotron emission
associated with an AGN. Flat spectrum sources ($-0.5<\alpha_r<0$) can
be produced by AGN activity or by optically thin Bremsstrahlung radiation from
star formation at higher radio frequencies ($\nu>5$\ GHz). Steep
spectrum sources ($\alpha_r<-0.5$) consist of diffuse synchrotron
emission often associated with either radio jets or
star formation in galaxies. 

In this paper we use the radio spectral 
index as a crude discriminator between AGN and star formation activity.
We arbitrarily classify any source with an available 
spectral index that has $\alpha_r>-0.3$ (there are six such sources in 
our sample) to be primarily powered by AGN activity.

\subsection{Optical, Near-infrared, and Mid-infrared Imaging}
\label{secimaging}

Wide-field and deep $HK'$ observations of the HDF and flanking fields
were obtained using the University of Hawaii Quick Infrared Camera
(QUIRC; \markcite{hodapp96}Hodapp et al.\ 1996) on the 2.2\ m 
University of Hawaii (UH) telescope
and the 3.6\ m Canada-France-Hawaii Telescope (CFHT).
The $HK'$ ($1.9\ \pm 0.4$\ $\mu$m) filter is described in
Wainscoat \& Cowie (in preparation); the empirical relation between
$HK'$ and $K$ is $HK'-K=0.13+0.05(I-K)$, which
simplifies to $HK'-K=0.3$, assuming the median $I-K$ galaxy
color (\markcite{barger99a}Barger et al.\ 1999a).
In Table~1 we use a $2\sigma$ limit of $HK'=21.5$ for our 
wide-field image 
and a $2\sigma$ limit of $HK'=22.6$ for our deep image in the
area around the HDF proper.

Deep Johnson $V$ and Kron-Cousins 
$I$-band observations were made with the CFHT over a much larger 
area using the UH8K CCD Mosaic Camera built by Metzger, Luppino, and Miyazaki.
Details of the above $HK'$ and optical observations can be found in
\markcite{barger99a}Barger et al.\ (1999a).
In Table~1 we use $2\sigma$ limits of $I=25.3$ and $V=26.4$.

In February 1998 we observed four of the sources in our radio sample 
with the near-infrared camera 
(NIRC; \markcite{matthews94}Matthews \& Soifer 1994) on the 
Keck~I 10\ m telescope. These sources previously had only $HK'$ limits
but were detected in the NIRC observations.
Conditions were photometric with seeing $\sim 0.7''$ FWHM.
NIRC has a $256\times 256$ InSb array with 
$0.15''\times 0.15''$ pixels, giving a $38''\times 38''$ field of view.
We imaged at 2.1\ $\mu$m ($K'$) with total exposure times for each object 
of $3240-4320$\ s. The data were obtained
in sets of 120\ s exposures, and
the center of the field was moved in a $3\times 3$ grid pattern with
$3''$ on a side. The centers of successive grids were moved by
$2''$ between each set. 
A fifth source was subsequently detected with NIRC with
a longer exposure under non-photometric conditions.
The data were processed using median sky flats
generated from the dithered images and calibrated onto the $HK'$
images using other galaxies in the field.

We used the Low-Resolution Imaging Spectrometer
(LRIS; \markcite{oke95}Oke et al.\ 1995) on the Keck~II 10\ m telescope
in March 1997 and February 1998 to obtain
$B$-band and Kron-Cousins $R$-band images, respectively, of a strip
region $6'\times 2.5'$ in size that crosses the HDF.
The total exposure times were 1680\ s and 1600\ s for $B$ and $R$, 
respectively, and the seeing was $0.8''$ in $R$ and $1.3''$ in $B$.
The $2\sigma$ limits are $B=26.6$ and $R=26.6$.
Details of the observations can be found in 
\markcite{cowie98}Cowie \& Hu (1998) and
\markcite{cowie99}Cowie, Songaila, \& Barger (1999).

Finally, a deep $U'(3400\pm 150$\AA) image of an area
$80$\ arcmin$^2$ centered on the HDF-proper was obtained using the
ORBIT CCD on the UH telescope. The $2\sigma$ limit for the $U'$ image
is $U'=25.8$. Details of the $U'$ observations can be found 
in Wilson et al.\ (in preparation). 

All photometric magnitudes were measured in
$3''$ diameter apertures and then corrected to $6''$ diameter (near total) 
magnitudes following the
procedures of \markcite{cowie94}Cowie et al.\ (1994).

\markcite{s97}Serjeant et al.\ (1997) obtained deep ISOCAM 
observations of the HDF and flanking fields at 6.75 and 15\ $\mu$m.
The \markcite{aussel99}Aussel et al.\ (1999) 
reductions of the data produced a main source list of
49 objects ($7\sigma$) and a supplementary list of an additional 
51 objects ($3\sigma$ for 15\ $\mu$m and $5\sigma$ for 6.75\ $\mu$m). 
Despite the large point spread function ($15''$ at 15\ $\mu$m), 
in most cases the
optical or NIR identification of the ISOCAM source was straightforward.
The redshifts for sources in the sample are in the range $z=0.078$ to 
$z=1.242$ (median $z=0.585$).

\subsection{Keck Spectroscopy}
\label{seckeck}

We used LRIS during a two night run on Keck~II in
March 1999 to obtain spectroscopic observations of a
sub-sample of the 1.4\ GHz sources in a strip region centered on the HDF.
Although the HDF and flanking field region has been intensely studied 
by a number of groups
(see \markcite{cohen99}Cohen et al.\ 1999 for a summary and references), 
most of the sources in our sample had not been previously observed due
to their faint optical fluxes.
We used $1.4''$ wide slits and the 400 lines\ mm$^{-1}$
grating blazed at 8500\ \AA, which gives a
wavelength resolution of $\sim 12$\,\AA\ and a wavelength coverage of
$\sim 4000$\ \AA. The wavelength range for each object depends on the 
exact location of the slit in the mask but is generally between
$\sim4000$ and 10000\ \AA. Three of the slit masks were
constructed at a position angle of $90^\circ$, and the remaining three
were nearly identical versions constructed
at a position angle of $-90^\circ$. 
This procedure enabled us to obtain sufficient wavelength
coverage for all the objects in our sample, including those that 
fell close to the edges of the masks.
The observations were 1.5\ hr per slit mask,
broken into three sets of 30\ minute exposures. Three HDF slit masks were
observed per night. Some of the objects were in all of the
slit masks and hence were observed for 9\ hrs (see Table~\ref{tab2}).
Conditions were photometric with seeing $\sim 0.6''-0.7''$ FWHM both 
nights. The objects were stepped along the slit by $10''$ in each
direction, and the sky backgrounds were removed using the median of 
the images to avoid the difficult and time-consuming problems of 
flat-fielding LRIS data. Details of the spectroscopic reduction 
procedures can be found in \markcite{cowie96}Cowie et al.\ (1996).

\begin{deluxetable}{ccccc}
\tablewidth{0pt}
\tablenum{2}
\tablecaption{1999 March Keck~II LRIS Spectroscopic Observations\label{tab2}}
\tablehead{
\colhead{\#} & \colhead{$HK'$} & \colhead{$I$} &
\colhead{Redshift} & \colhead{Exposure (hr)}
}
\startdata

16 & 23.39 & 25.30 & \nodata & 6 \\
17 & 18.75 & 22.49 & 1.013 & 9 \\
23 & 16.36 & 19.09 & 0.456 & \nodata \\
24 & 19.10 & 22.29 & 1.219 & \nodata \\
25 & 19.25 & 22.71 & \nodata & 3 \\
26 & 22.29 & 25.30 & \nodata & 6 \\
28 & 21.22 & 25.30 & \nodata & 6 \\
30 & 21.23 & 25.30 & \nodata & 3 \\
31 & 18.06 & 20.86 & 0.857 & \nodata \\
32 & 17.85 & 21.04 & 1.013 & \nodata \\
33 & 21.65 & 24.72 & \nodata & 3 \\
35 & 18.36 & 20.76 & 0.961 & \nodata \\
36 & 22.60 & 25.30 & \nodata & 3 \\
37 & 21.00 & 23.20 & \nodata & 3 \\
38 & 18.74 & 21.09 & 0.475 & \nodata \\
40 & 17.36 & 19.88 & 0.410 & \nodata \\
41 & 21.50 & 25.30 & \nodata & \nodata \\
42 & 16.52 & 18.70 & 0.321 & \nodata \\
43 & 19.31 & 21.88 & 1.275 & \nodata \\
44 & 21.50 & 25.30 & \nodata & 6 \\
46 & 22.60 & 25.30 & \nodata & 9 \\
47 & 19.55 & 22.64 & 0.474 & \nodata \\
48 & 18.52 & 20.92 & 0.761 & \nodata \\
50 & 20.14 & 24.81 & 0.884 & \nodata \\
51 & 19.21 & 21.89 & 1.243 & \nodata \\
52 & 18.18 & 20.55 & 0.902 & \nodata \\
53 & 20.29 & 24.70 & \nodata & 3 \\
54 & 20.76 & 25.13 & \nodata & 3 \\
56 & 17.98 & 19.87 & 0.422 & \nodata \\
59 & 19.51 & 22.66 & \nodata & 3 \\
60 & 20.08 & 22.90 & \nodata & 6 \\
61 & 20.38 & 22.90 & \nodata & \nodata \\
62 & 17.36 & 19.80 & 0.558 & \nodata \\
64 & 17.59 & 19.90 & 0.411 & 7.5 \\
65 & 20.90 & 25.30 & \nodata & 3 \\
66 & 19.39 & 22.14 & 1.019 & 3 \\
69 & 21.50 & 25.30 & \nodata & 1.5 \\
\enddata
\end{deluxetable}

\begin{figure*}[tb]
\centerline{\psfig{figure=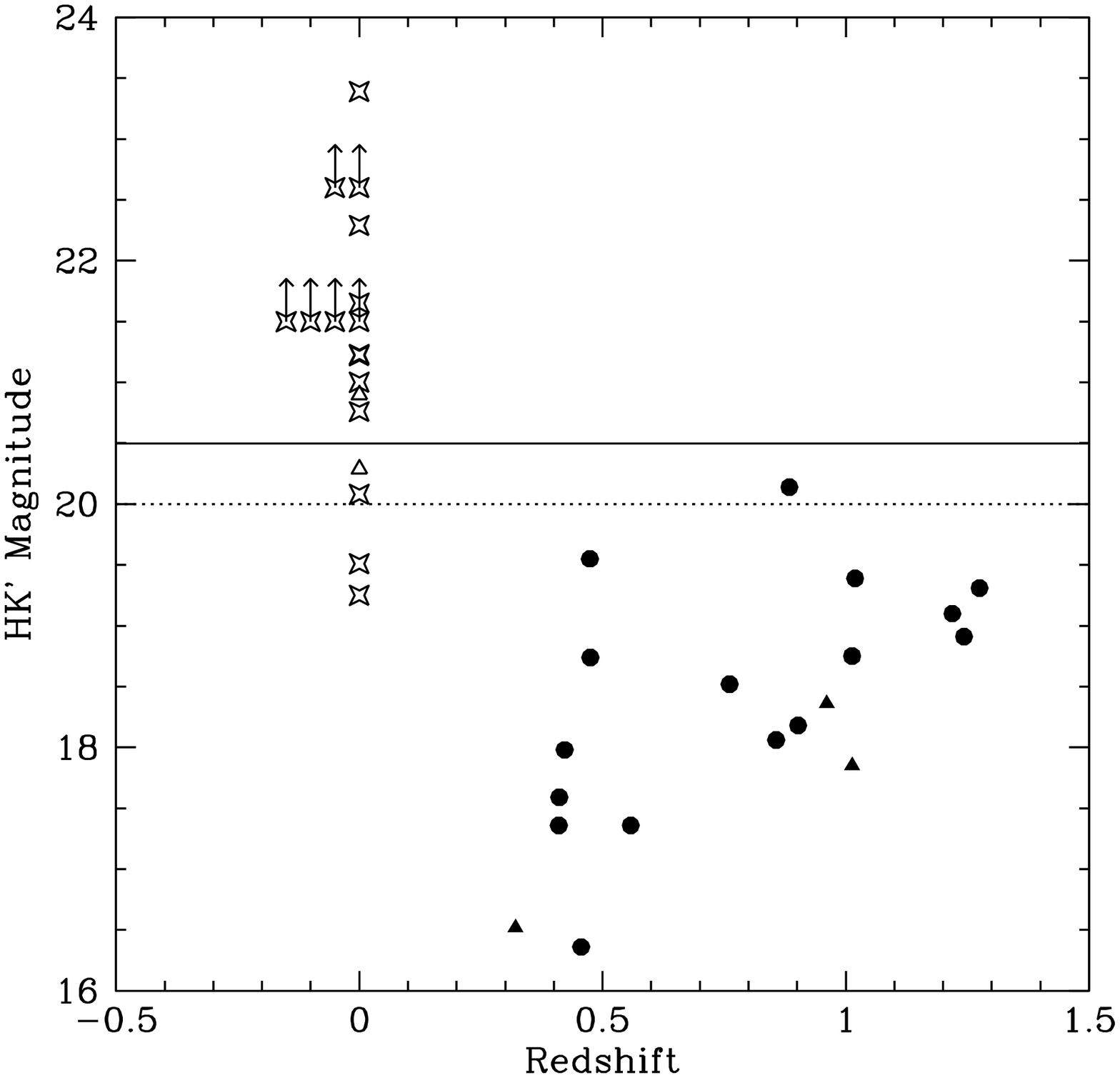,width=3.5in}
\psfig{figure=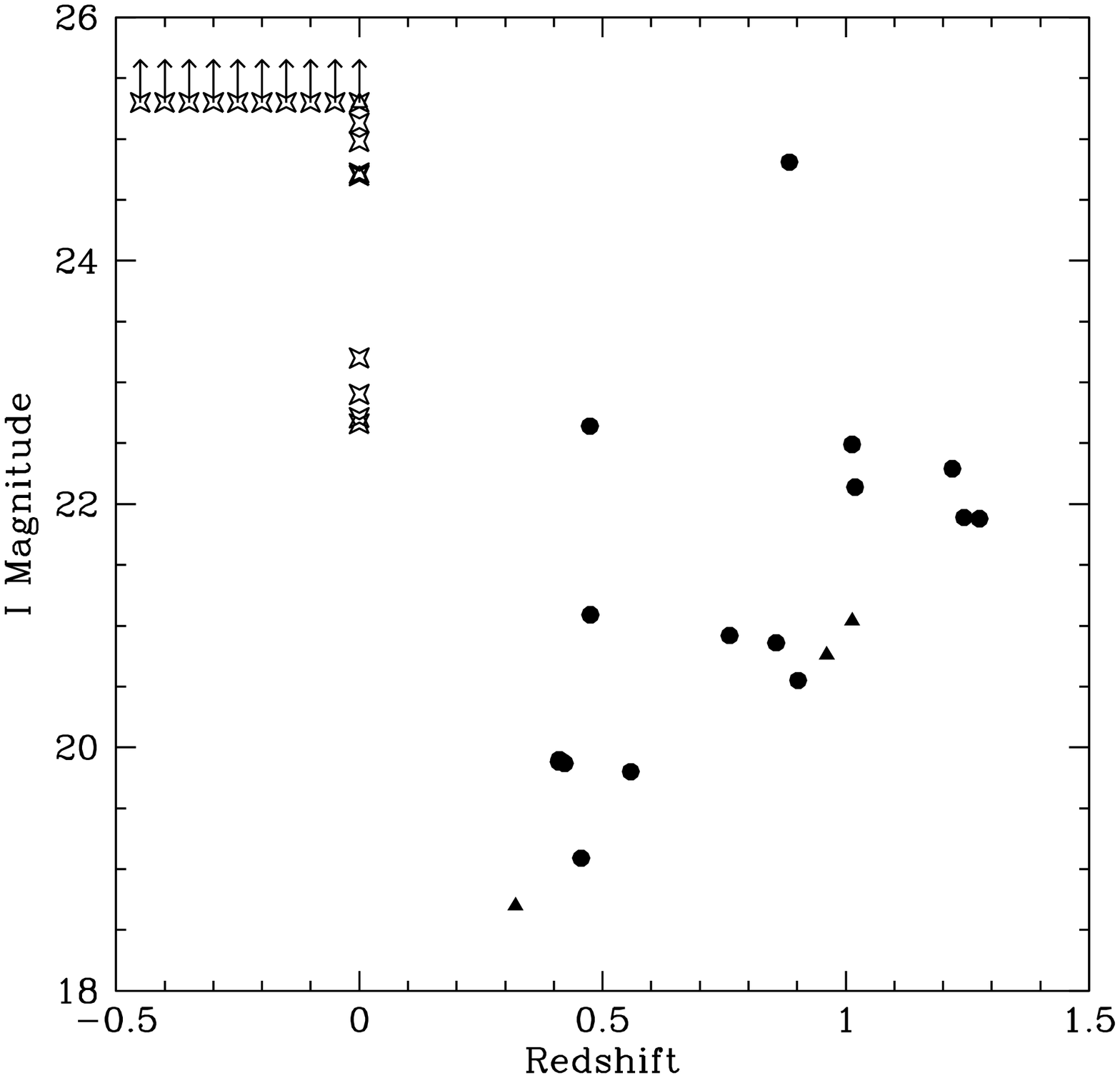,width=3.5in}}
\figurenum{1}
\figcaption[]{
(a)\ $HK'$ and (b)\ $I$ versus redshift for our
LRIS sample. Magnitude limits are indicated by arrows.
Open symbols at $z\lesssim 0$ do not have redshifts.
Triangles have spectral indices $>-0.3$, indicating AGN.
Dotted line in (a) at $HK'=20$ is the practical spectroscopic
limit. Solid line at $HK'=20.5$ is our bright magnitude bound
to the optical/NIR-faint radio sample (see \S 2.3).
\label{figzhkandzi}
}
\end{figure*}

In Table~\ref{tab2} we list all the objects in the radio sample that fall in
the LRIS strip region. The columns are radio catalog number from 
Table~1,
$HK'$\ mag, $I$\ mag, redshift, and exposure time for any object
targeted in our March 1999 spectroscopic survey. 
Over an area $\sim 58$\ arcmin$^2$, 19 of the 37 radio sources now 
have secure redshift identifications; all are at $z\lesssim 1.3$.
Figure~\ref{figzhkandzi}a, b shows $HK'$ versus redshift
and $I$ versus redshift. Spectroscopic redshifts are in general 
relatively straightforward to obtain for radio objects with 
$HK'\lesssim 20$.

We note that Waddington et al.\ (1999) claim a redshift
identification of $z=4.42$ for object 30 based on a Ly$\alpha$ detection.
However, the position of their detection is $1''$ away from the radio
source position and counterpart optical galaxy, and so an association
with the radio source is not secure. The radio
source is detected in the mid-IR, which would suggest $z<1.3$, consistent
with the redshift estimated using the millimetric redshift 
technique described in \S~\ref{secze}. However, we did not see any strong
[O\,II]\ 3727 feature in our spectrum
from $4000-9200$\ \AA, which would suggest $z>1.5$.
Thus, the redshift of this object remains uncertain.

%
%

In all of our slit-masks (equivalent to a 9\ hr integration)
we also included the position of the brightest SCUBA source, HDF850.1,
from the deep 850\ $\mu$m map of the HDF-proper by H98.
We centered our slit across the
position of the 1.3\ mm detection of HDF850.1 reported by
Downes et al.\ (1999), which also coincides with the 8.5\ GHz supplemental
radio source 3651+1226 (Richards et al.\ 1998). We oriented the slit such
that it fell across both the arc-like feature 3-593.0 that is favored
by Downes et al.\ (1999) as the optical counterpart to HDF850.1
and the nearby red galaxy 3-586.0.
However, we were unable to determine a secure redshift for
either source from our spectroscopic data.

\subsection{Submillimeter Observations}
\label{secsmm}

Our SCUBA jiggle map observations were flexibly scheduled in mostly 
excellent observing conditions during 
two runs in April and June 1999 for a total of five observing shifts. 
The maps were dithered to prevent any regions of
the sky from repeatedly falling on bad bolometers. The chop throw was
fixed at a position angle of $90^\circ$ so that the negative beams would
appear 45\ arcsec on either side east-west of the positive beam. The
data were reduced using beam weighted extraction routines that included 
both the positive and negative portions of the chopped images,
thereby increasing the effective exposure times. Regular ``skydips''
(\markcite{manual}Lightfoot et al.\ 1998) were obtained to measure the 
zenith atmospheric opacities at 450 and 850\ $\mu$m, and the 225\ GHz sky
opacity was monitored at all times to check for sky stability.
The median 850\ $\mu$m optical depth for all nights together was 0.265.
Pointing checks were performed every hour during the observations on
the blazars 0954+685, 1418+546, 0923+392, or 1308+326. The data were
calibrated using 30\ arcsec diameter aperture measurements of the positive
beam in beam maps of the primary calibration source, Mars, and one of
three secondary calibration sources, CRL618, IRC+10216, or OH231.8.

%
%

\begin{figure*}[tb]
\figurenum{2}
\centerline{\psfig{figure=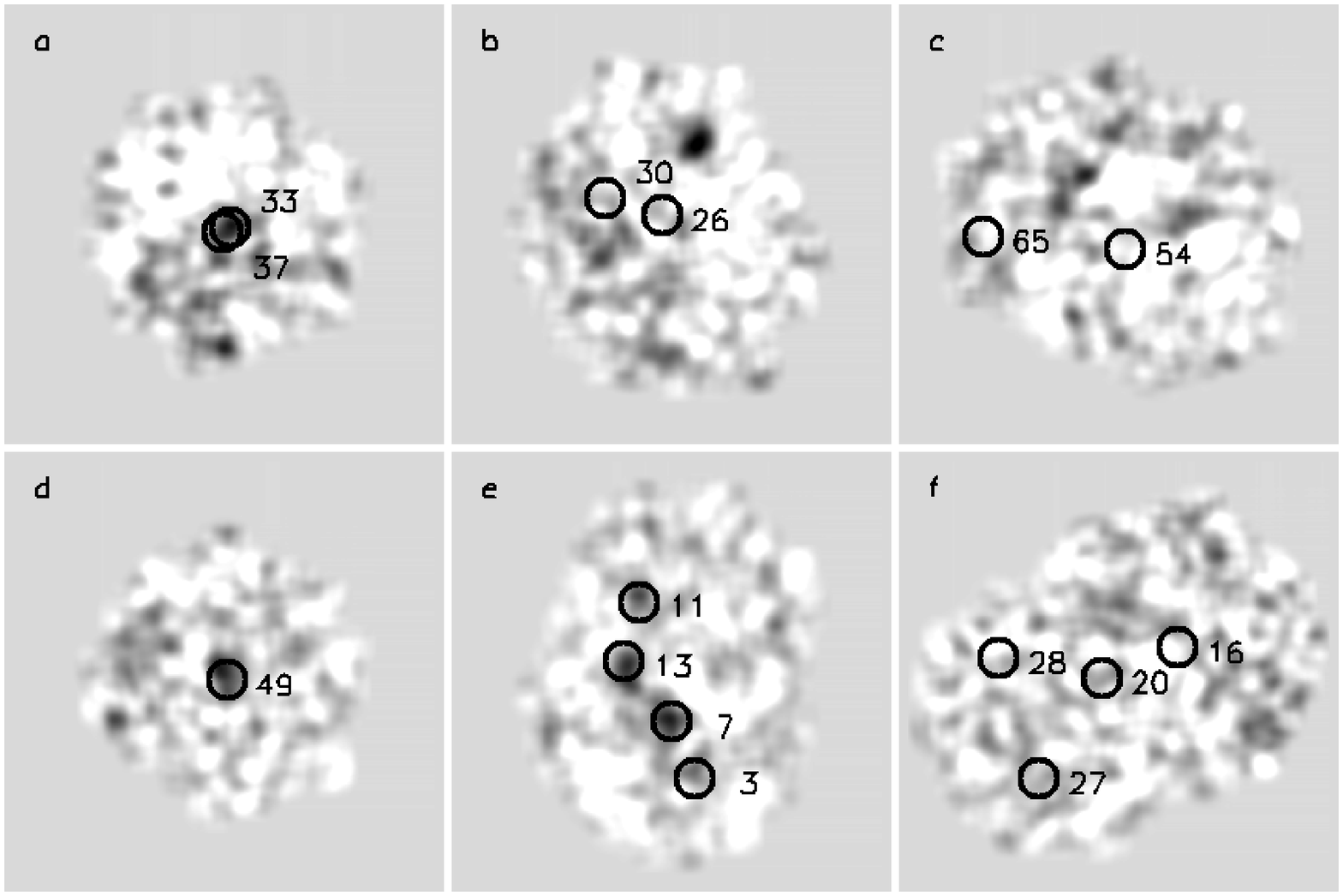,angle=90,width=6.2in}}
\figcaption[]{
850\ $\mu$m SCUBA maps of our 6 radio-selected fields
(excluding the H98 map) overlayed with circles at the positions of all
optical/NIR-faint 1.4\ GHz sources. North is up and East is to the left
in the images. Labels are the catalog numbers from Table~1
for the optical/NIR-faint sources.
\label{figmaps}
}
\end{figure*}

The data were reduced in a standard and consistent way using the 
dedicated SCUBA User Reduction Facility 
(SURF; \markcite{surf}Jenness \& Lightfoot 1998).
Due to the variation in the density of bolometer samples across the maps,
there is a rapid increase in the noise levels at the very edges.
We have clipped the low exposure edges from our images.
We present our SCUBA maps in Fig.~\ref{figmaps}.

We also re-reduced the archival HDF-proper SCUBA data of H98
in order to make a consistent analysis with the present data.
We could only make use of the maps that were
taken with a fixed RA chop (90\ per cent of the data sample), 
which included 34 hours with a 30\ arcsec chop throw and 28\ hours with 
a 45\ arcsec chop throw. We combined these data separately to form 
two independent maps. In Table~1 we quote 
submillimeter fluxes determined from the weighted average of
measurements made in each map.

The SURF reduction routines arbitrarily normalize all the data
maps in a reduction sequence to the central pixel of the first 
map; thus, the noise levels in a combined image are 
determined relative to the quality of the central pixel in the
first map. In order to determine the absolute noise levels of
our maps, we first eliminated the $\gtrsim 3\sigma$ real sources in
each field by subtracting an appropriately normalized version
of the beam profile. We then iteratively adjusted the noise
normalization until the dispersion of the signal-to-noise
ratio measured at random positions became $\sim 1$. 
Our noise estimate includes both fainter sources and correlated noise.

%
%

\begin{figure*}[tb]
\figurenum{3}
\centerline{\psfig{figure=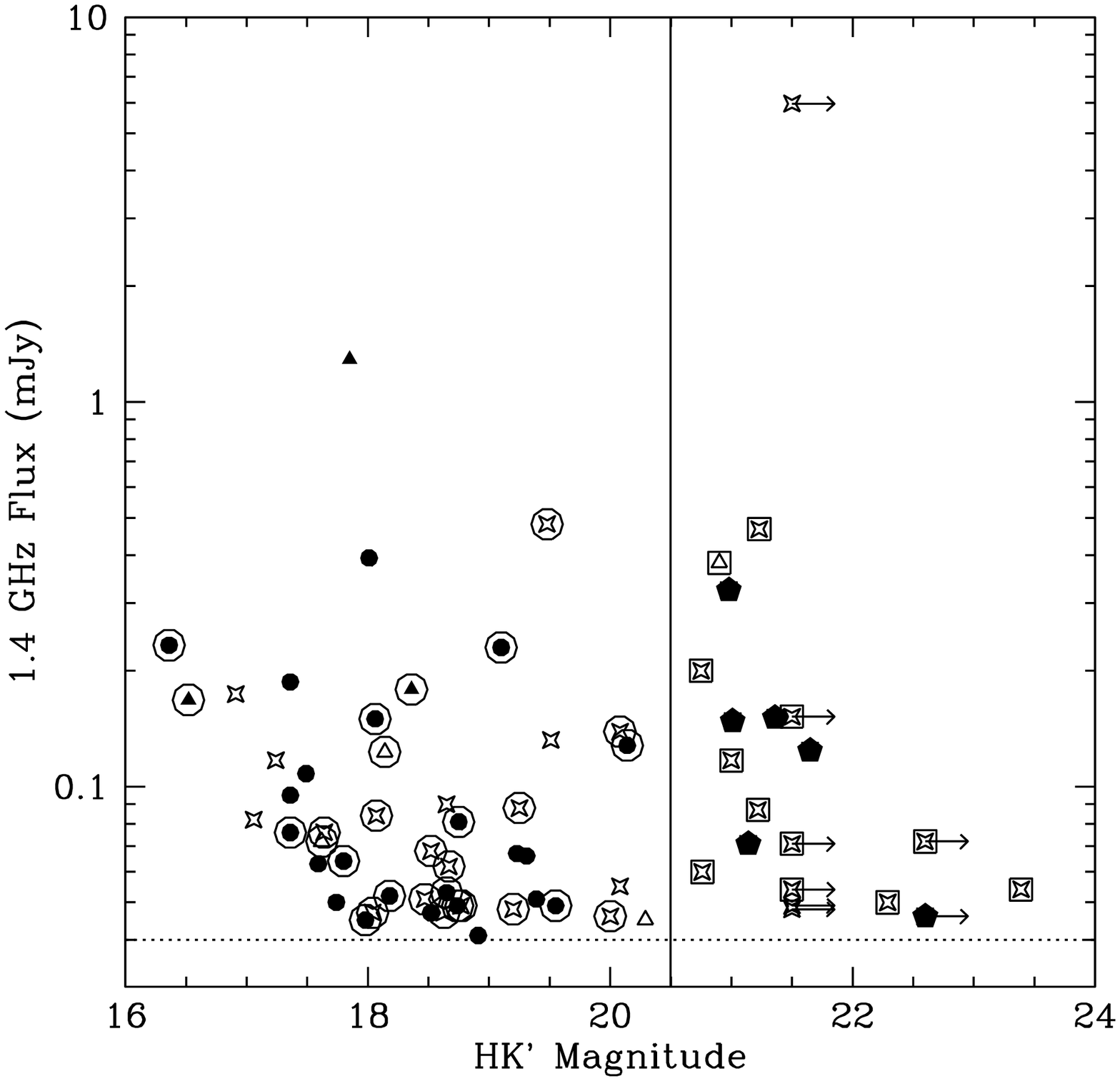,width=3.5in}
\psfig{figure=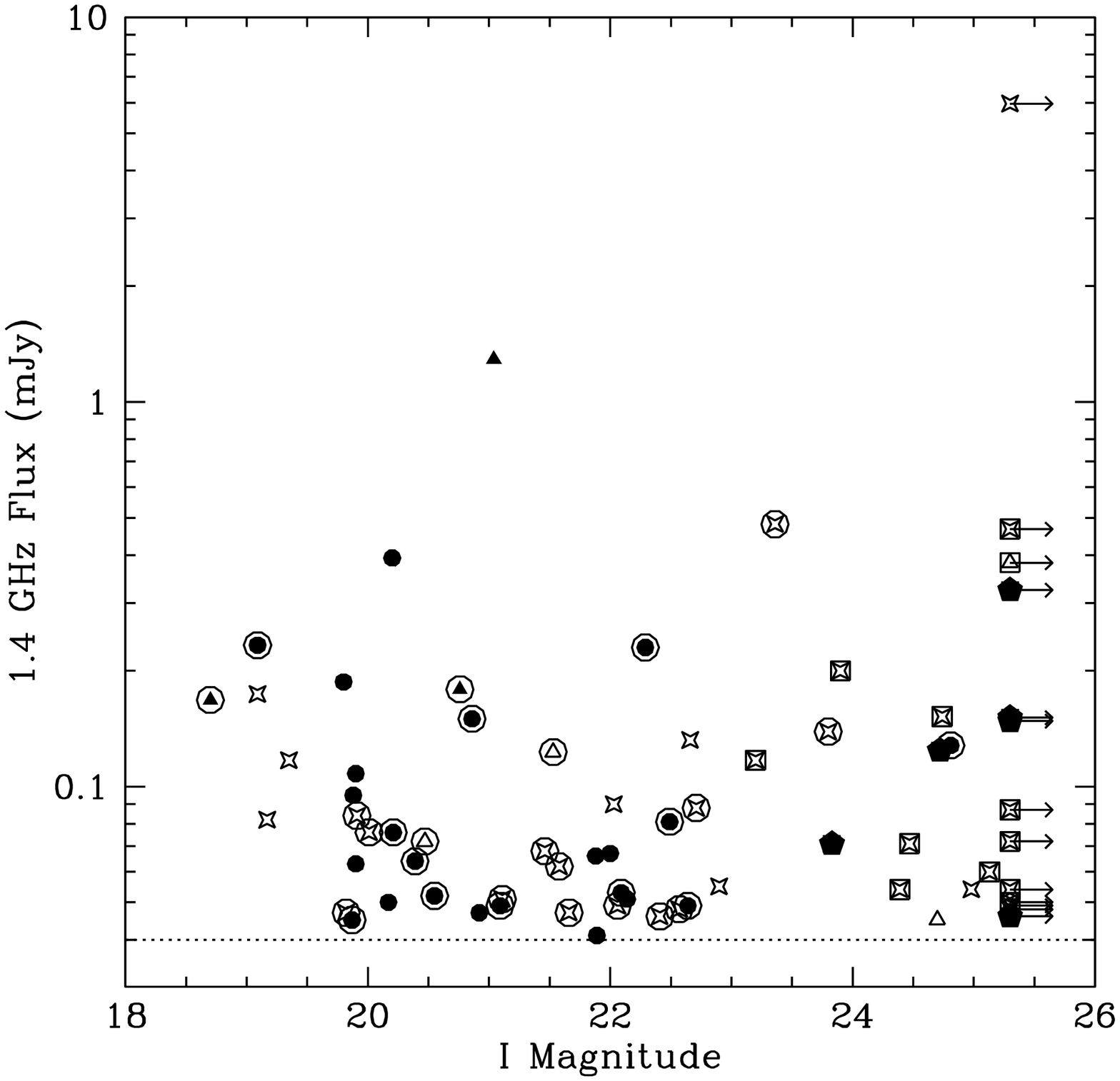,width=3.5in}}
\figcaption[]{
1.4\ GHz flux versus (a)\ $HK'$ and (b)\ $I$\ mag for our
radio sample. Filled circles have known redshifts (all are
at $z<1.3$); open crosses do not.
Filled (known redshifts) and open triangles have $\alpha_r>-0.3$.
Open circles were observed at 353\ GHz (850\ $\mu$m)
but not detected at the typical 6\ mJy
($3\sigma$) level. Filled pentagons have submillimeter detections.
Dotted horizontal line is the 1.4\ GHz flux limit of 0.040\ mJy.
Vertical line in (a) is our $HK'=20.5$ optical/NIR-faint radio
source bright magnitude limit.
\label{figcolor}
}
\end{figure*}

We centered on the positions of the radio sources and measured the
submillimeter fluxes in both the positive and negative beams.
The extracted fluxes were calibrated to the 30\ arcsec diameter
aperture fluxes of the brightest sources.
The brighter submillimeter sources were subtracted from the
maps before the fainter sources were extracted. In cases where
the positions of two radio sources are very close, we may be
slightly overestimating the submillimeter flux allocated to
the bright source with this procedure
and underestimating that allocated to the faint source,
but since the total submillimeter flux should be reasonable, we will
not be making any gross errors in our later estimates of the total
star formation rate.

From the total sample of 70 radio-selected galaxies in the
$\sim 80$\ square arcminute central region of the radio map,
we take the 22 with $HK'>20.5$ to be
our optical/NIR-faint radio sample, for which there are
now submillimeter observations of 19.
Even though our SCUBA observations were relatively shallow
(the 850\ $\mu$m $3\sigma$ limit was 6\ mJy), we detected 5 of the 15
optical/NIR-faint radio sources that we observed;
a sixth significant source ($S_{850\ \mu{\rm m}}=2.4\pm 0.7$\ mJy)
was detected in the HDF-proper SCUBA map of H98
(HDF850.2 in their notation). The submillimeter flux we
measure for the brightest source (HDF850.1) in the
HDF-proper map is $5.4\pm 0.6$\ mJy, which is slightly lower
than the value of $7.0\pm 0.4$\ found by H98 but is
consistent within the statistical and systematic errors.
The jiggle map observing mode enabled simultaneous
observations of a large fraction (31/48) of the optical/NIR-bright radio
sources, none of which were detected.
There are two $>6$\ mJy submillimeter sources in our jiggle
maps that were not in the radio sample.

In Fig.~\ref{figcolor}a, b we show the 1.4\ GHz fluxes of the
radio sources in our sample versus their $HK'$ and $I$ magnitudes,
indicating those with measured redshifts and those with submillimeter
detections.

\section{Submillimeter Source Counts}
\label{secsmmcounts}

\begin{figure*}[tb]
\figurenum{4}
\centerline{\psfig{figure=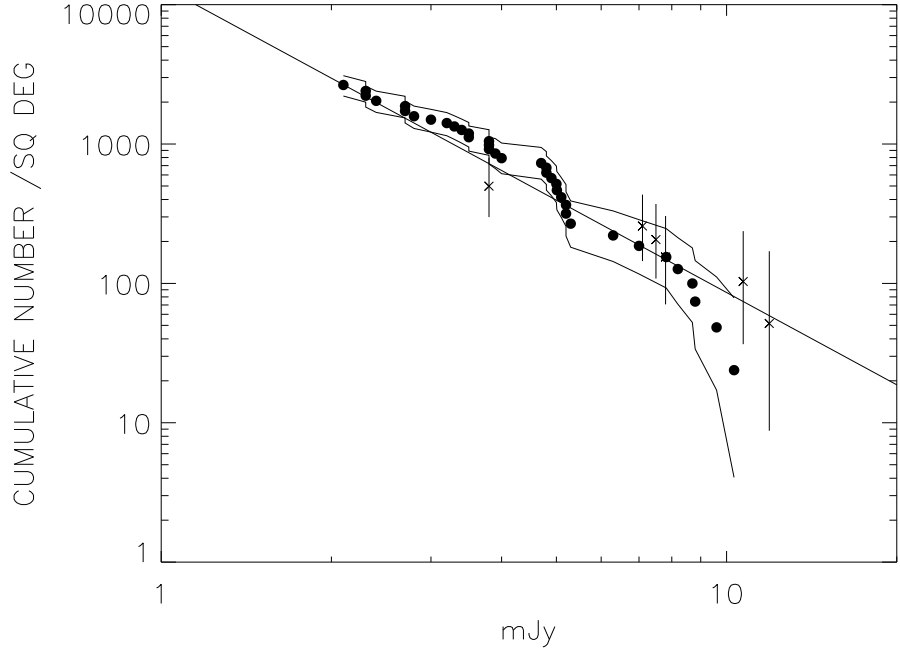,angle=90,width=5.0in}}
\figcaption[]{
Comparison of the radio-selected 850\ $\mu$m source counts (X symbols with
$1\sigma$ uncertainties) with the combined counts
(solid circles) from blank field submillimeter surveys
(H98; Eales et al.\ 1999; Barger, Cowie, \& Sanders 1999).
Jagged solid lines are $1\sigma$ uncertainties.
Solid line is a power-law parameterization, $N(>S)\propto S^{-2.2}$
(Barger, Cowie, \& Sanders 1999). Radio selection detects the majority
of the bright submillimeter source population.
\label{figcounts1}
}
\end{figure*}

We find that the radio selection technique is effective in locating
the majority of bright submillimeter sources.
We document this in Fig.~\ref{figcounts1} where we compare the combined 
850\ $\mu$m source counts from blank field submillimeter surveys
(H98; Eales et al.\ 1999; Barger, Cowie, \& Sanders 1999)
with our new radio-selected 850\ $\mu$m source 
counts. A source of flux strength $S_i$ contributes $N(S_i)=1/A_i$ to
the counts per unit area, where $A_i$ is the area over which
there is $\ge 3\sigma$ sensitivity to $S_i$. The cumulative counts,
$N(>S)$, are given by the sum of the inverse areas of all sources
brighter than $S$. However, for our radio-targeted search, the
effective area (65 square arcminutes) is greater than if we had done 
an untargeted search with the same number of pointings 
and is given by the fraction (number $HK'>20.5$ sources 
observed)/(total number $HK'>20.5$ sources in sample) times the
radio survey area (79.4\ square arcminutes).
Because of the correspondence between
the optical/NIR-faint radio population and the bright submillimeter
population, we do not need to survey the entire radio field in order to
observe the bright submillimeter population.

It is interesting to speculate whether the correspondence
between optical/NIR-faint radio sources and bright submillimeter
sources also holds at fainter submillimeter flux levels. A necessary
requirement for this to be the case is that the surface density
of faint radio sources be comparable to or exceed the surface
density of submillimeter sources. Extrapolating a power-law
parameterization of the 1.4\ GHz counts over the range $40-1000$\ $\mu$Jy
(Richards 1999b) down to a 1\ $\mu$Jy cut-off gives 
$N(S_{1.4}>1\ \mu$Jy) $=127$\ arcmin$^{-2}$. 
The empirically estimated surface density of submillimeter sources
(Barger, Cowie, \& Sanders 1999) is $N(S_{353}>0)\simeq 11$\ arcmin$^{-2}$.

A possible counter-argument to a close correspondence between
radio and submillimeter sources at faint submillimeter flux levels comes
from the study by \markcite{smail99b}Smail et al.\ (1999b) who compared
their lensed submillimeter survey data with radio data.
They found that the Carilli \& Yun (1999)
radio-submillimeter spectral indices differed for their bright and faint
submillimeter subsamples. They suggested that the difference
could arise if the faint submillimeter sources lie at higher redshifts
or have different fractions of radio-loud
AGN or have different dust temperatures. However, with the low statistics,
strong conclusions cannot be drawn at this time.

%
%

\section{Millimetric Redshift Estimation}
\label{secze}

Optical spectroscopic surveys of submillimeter galaxies are difficult
because of the very different behaviors of the $K$-corrections
in the optical and submillimeter and the poor submillimeter resolution.
Spectroscopic surveys to date
(\markcite{barger99b}Barger et al.\ 1999b;
\markcite{lilly99}Lilly et al.\ 1999) 
give potentially conflicting results and are very limited in size.
Photometric redshift estimates of submillimeter sources have been
made from candidate optical counterparts (H98;
Lilly et al.\ 1999), but these are questionable both because the
counterpart identification is not secure and because the
SEDs of the submillimeter sources may be unlike those of the optical 
sources used in making the redshift estimates due to dust extinction
and possible AGN contributions.
Consequently, a new approach is required if the
positions of the submillimeter sources are to be reliably determined
and their nature and redshift distribution understood.

%
%

\begin{figure*}[tb]
\figurenum{5}
\centerline{\psfig{figure=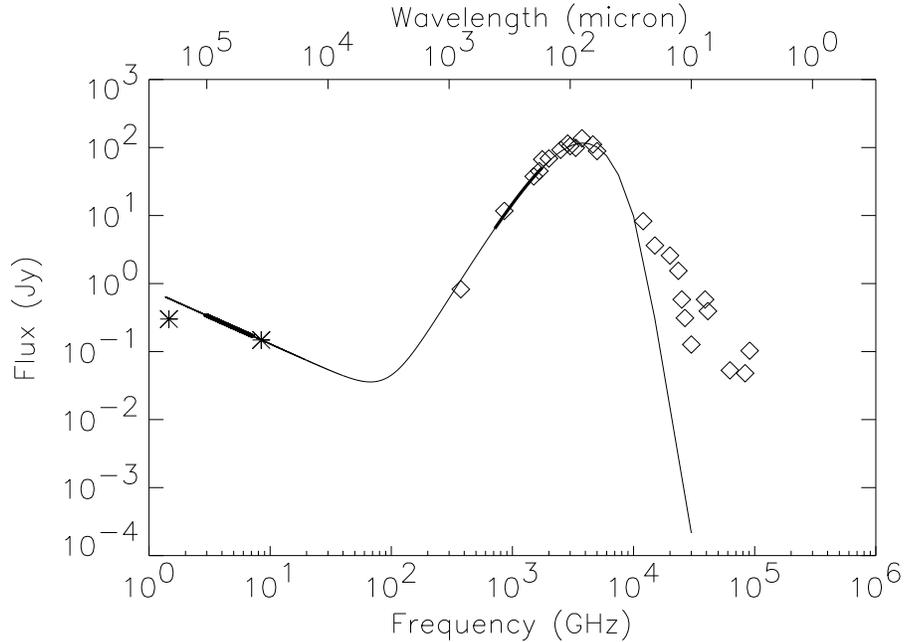,angle=90,width=5.0in}}
\figcaption[]{
Open diamonds are the Arp~220 FIR/submillimeter flux values
from Klaas et al.\ (1997) and
Rigopoulou et al.\ (1996); asterixes
are the 1.4\ GHz and 8.4\ GHz points from
Condon et al.\ (1991). Smooth curve is from Eqs.~3 and 5.
Heavy lines indicate the
regions of the SED that are sampled over the redshift range
$z=1-4$ and illustrate the opposing spectral slopes.
\label{figarp}
}
\end{figure*}

The remarkably tight local correlation between the global FIR and
nonthermal radio luminosities provides a promising alternative
method for identifying and studying individual submillimeter sources,
provided that the correlation holds to high redshift.
In \S~\ref{secsmmcounts} we found that targeting with SCUBA 
optical/NIR-faint 1.4\ GHz sources is an efficient technique for 
identifying the majority of the bright submillimeter source 
population. Our results indicate that a large fraction of bright sources 
in submillimeter surveys have extremely faint optical/NIR
counterparts and hence are inaccessible to optical spectroscopy
(see also \markcite{smail99b}Smail et al.\ 1999b).
This conclusion is consistent with results from the
\markcite{barger99b}Barger et al.\ (1999b)
spectroscopic survey of lensed submillimeter sources
discussed in the introduction.

\begin{figure*}[tb]
\figurenum{6}
\centerline{\psfig{figure=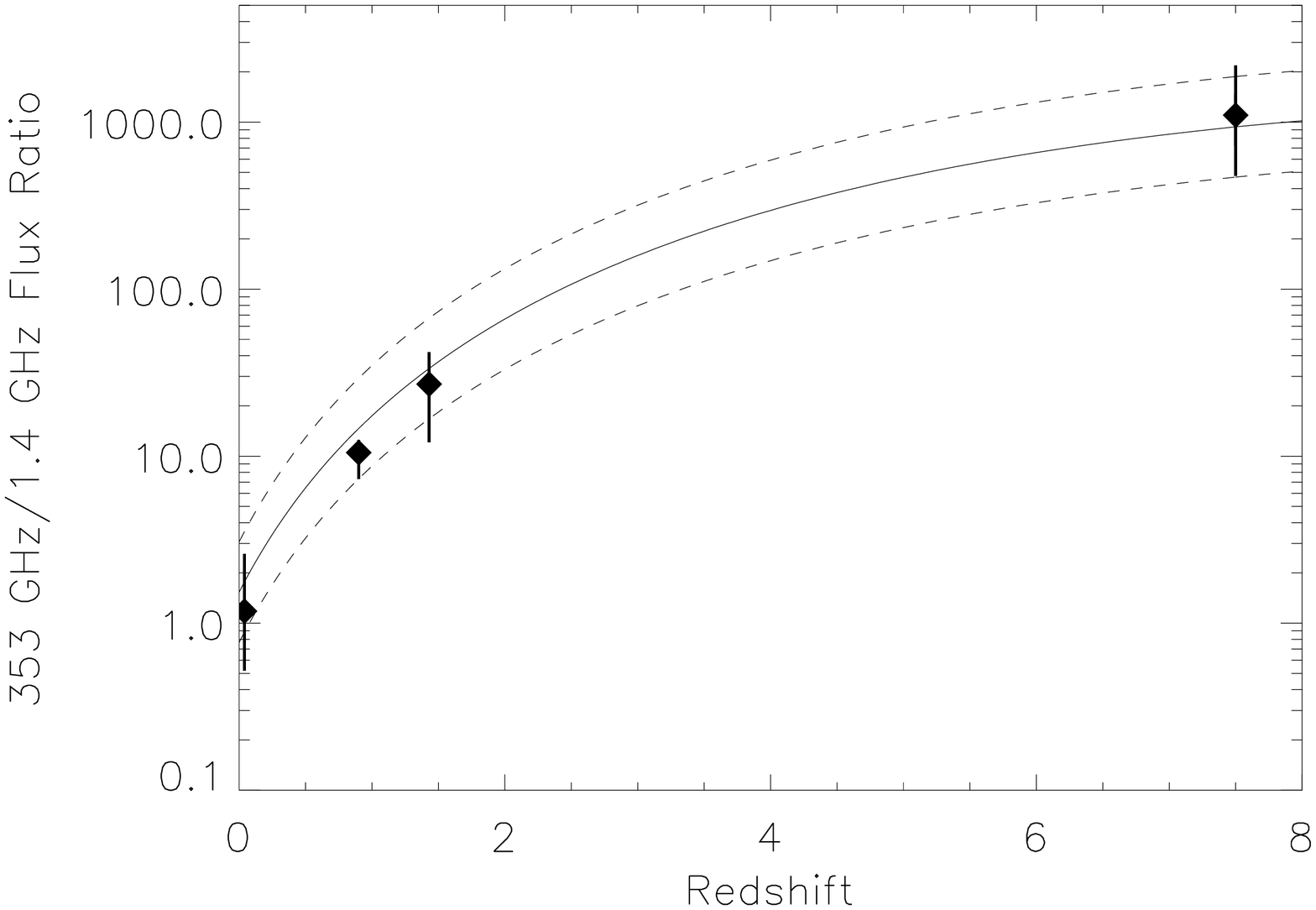,angle=90,width=3.8in}
\psfig{figure=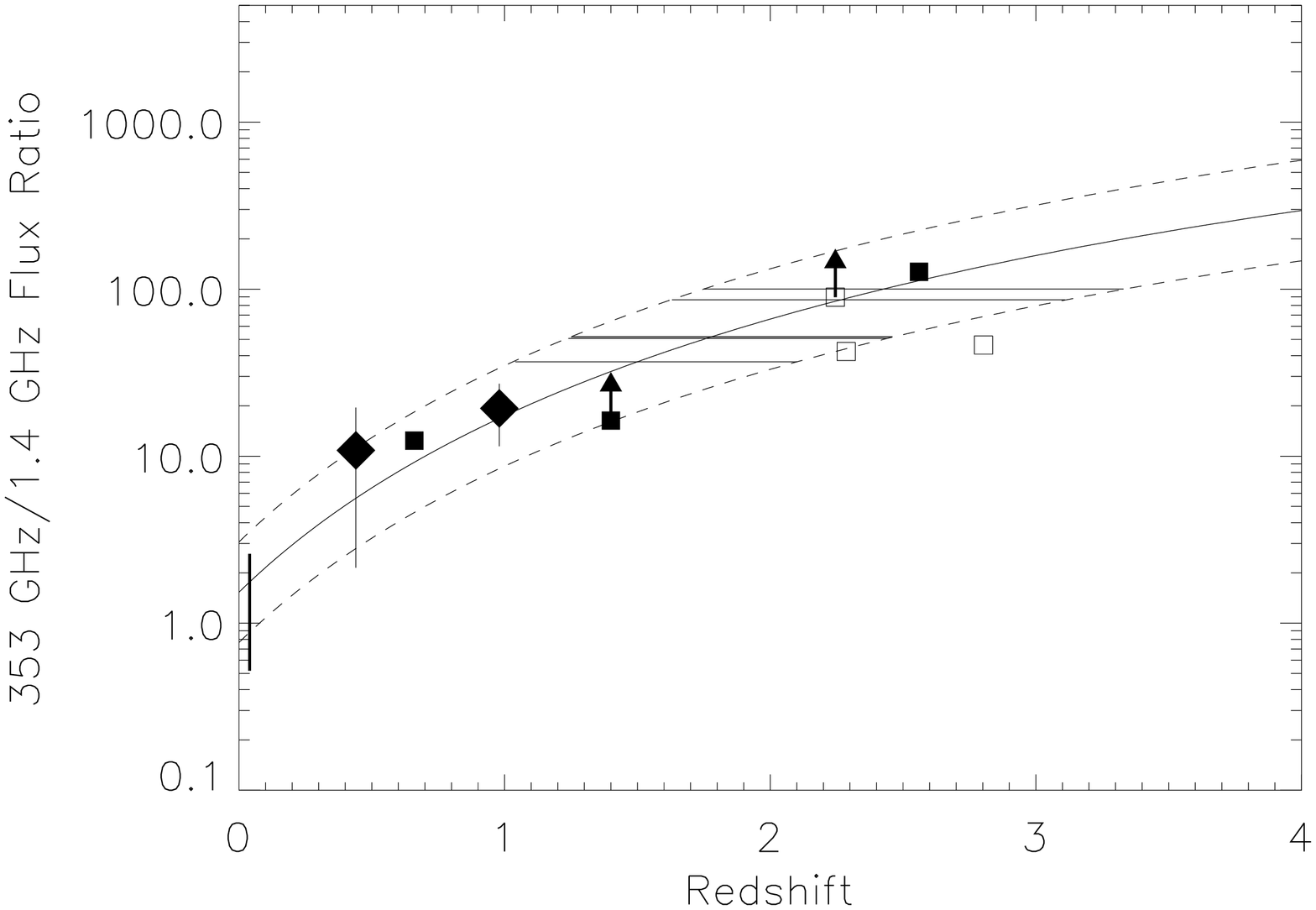,angle=90,width=3.8in}}
\figcaption[]{
(a)\ Solid curve is the 353\ GHz to 1.4\ GHz flux ratio
versus redshift for a non-evolving redshifted Arp~220.
Solid diamonds are the mean values of local ULIG data measured at
3000\ GHz (100\ $\mu$m), 857\ GHz (350\ $\mu$m),
667\ GHz (450\ $\mu$m), and 375\ GHz (800\ $\mu$m)
by Rigopoulou et al.\ (1996) and at 8.4\ GHz by Condon et al.\ (1991).
Sources are assumed to be at redshifts 7.5, 1.4, 0.9, and 0.06
to have been observed at 353\ GHz (850\ $\mu$m).
8.4\ GHz data were extrapolated to 1.4\ GHz at the above redshifts
by assuming a synchrotron spectrum with $\alpha_r=-0.8$.
Vertical ranges on the measurements are the minimum and maximum
flux ratios; the spread is about a multiplicative factor of two, as
indicated by the dashed curves.
(b)\ Curves and vertical bar at low redshift are as in (a).
Solid squares are individually detected star forming objects from
Lilly et al.\ (1999; CFRS 14 at $z=0.66$),
Dey et al.\ (1999; HR10 at $z=1.44$), and
Barger et al.\ (1999b; J1/J2 at $z=2.56$).
Open squares are AGN
from Rowan-Robinson et al.\ (1993; IRAS F10214+4724 at
$z=2.29$), Ivison et al.\ (1998; L1/L2 at $z=2.81$), and
Ivison et al.\ (1999; J6/J7 at $z=2.24$).
Large filled diamonds with $1\sigma$ uncertainties are
average submillimeter-to-radio flux ratios for two
redshift bins ($z\le 0.7$ and $0.7<z<1.3$).
Short horizontal lines are the redshift ranges
for our 6 significant radio-selected submillimeter detections.
\label{figulig}
}
\end{figure*}

Although we are unable to obtain spectroscopic
redshifts for the optical/NIR-faint radio-selected submillimeter
sources, we can use the submillimeter-to-radio flux ratios
to obtain millimetric redshift estimates.
Figure~\ref{figarp} illustrates how
the slope of a dusty galaxy's SED changes abruptly at frequencies 
higher than 100\ GHz. Below 30\ GHz synchrotron radio emission is
dominant, free-free emission is largest in the range $30-200$\ GHz, 
and thermal dust emission dominates above 200\ GHz.
Because of the opposing spectral
slopes of the blackbody spectrum in the submillimeter and the
synchrotron spectrum in the radio,
the submillimeter-to-radio flux ratio rises extremely rapidly with 
increasing redshift (see Fig.~\ref{figulig}).
Carilli \& Yun (1999; hereafter CY99) have therefore suggested using the
submillimeter-to-radio flux ratio as a redshift estimator.

%
%

The dust emission from local luminous infrared galaxies is well described
by optically thin single-temperature modified blackbodies with extinction
coefficient $\epsilon_\nu\propto\nu^\beta$, where $\beta \approx 1$ to 2.
The dust temperatures derived lie between 30 and 60\ K.
Arp~220, with its high bolometric luminosity produced almost
entirely from starburst activity (\markcite{ds98}Downes \& Solomon 1998), 
is an appropriate prototype for high redshift submillimeter sources,
as we will justify subsequently. Over the interval
$100\ {\rm GHz}\lesssim \nu\lesssim 10^{4}$\ GHz
($3000\ \mu {\rm m}\gtrsim \lambda\gtrsim 30\ \mu$m)
Arp~220's SED is well represented by a modified blackbody with
$\beta=1$ and a dust temperature $T_d=47$\ K; the luminosity
of this dominant cooler component is 
$1.36\times 10^{12}\ h_{65}\ {\rm L}_\odot$
(\markcite{klaas97}Klaas et al.\ 1997).
In terms of the blackbody distribution

\begin{equation}
B(\nu,T)={S(\nu,T)\over \pi}={{2h\nu^3}\over {c^2}}{1\over {e^{h\nu/kT}-1}}
\label{bb}
\end{equation}

\noindent
the submillimeter flux for an Arp~220-like galaxy at
redshift $z$ is given by

\begin{equation}
S_{s}= f (0.57\ {\rm Jy}) \pi B(\nu (1+z),T_d) \nu^\beta (1+z)^{1 + \beta}
\Biggl[{d_L(z_{Arp}) \over d_L(z)}\Biggr]^2
\label{smmexact}
\end{equation}

\noindent
where $\nu$ is the observed submillimeter frequency, $T_d$ is the 
dust temperature, and $d_L$ is the luminosity distance. 
We allow for an overall strength factor, $f$, relative 
to Arp~220, for which $f=1$.
The above normalization constant is fixed by a $\chi^2$ fit to the
Arp~220 flux measurements.
In the Rayleigh-Jeans long-wavelength limit, the submillimeter flux is
approximately given by
\begin{equation}
S_{s}= f (0.825\ {\rm Jy}) (1+z)^{1+\alpha_{s}} 
\Biggl({T_d\over 47\ {\rm K}}\Biggr)
\Biggl({\nu \over 375\ {\rm GHz}}\Biggr)^{\alpha_{s}}
\Biggl[{{d_L(z_{Arp})} \over {d_L(z)}}\Biggr]^2
\label{smmlong}
\end{equation}

\noindent
which is normalized to
the measured 375\ GHz (800$\mu$m) flux value for Arp~220 from
\markcite{rig96}Rigopoulou, Lawrence, \& Rowan-Robinson (1996);
here $\alpha_{s}=\beta+2=3$.
The approximation in Eq.~\ref{smmlong} is
reasonably accurate for $z\lesssim3$; at higher $z$ it extrapolates
above the result of Eq.~\ref{smmexact}.

The radio flux is given by

\begin{equation}
S_{r}=f (0.148\ {\rm Jy}) (1+z)^{1+\alpha_{r}}
\Biggl({\nu \over 8.4\ {\rm GHz}}\Biggr)^{\alpha_{r}}
\Biggl[{{d_L(z_{Arp})} \over {d_L(z)}}\Biggr]^2
\label{radio}
\end{equation}

\noindent
which is normalized to the measured 8.4\ GHz flux for Arp~220 from
\markcite{condon91}Condon et al.\ (1991). 
Since both $S_s$ and $S_r$ depend linearly on the
star formation rate, the scale factor $f$ is the same.
In local star forming galaxies there is an observed spectral 
flattening at 1.4\ GHz due to free-free absorption 
(\markcite{condon92}Condon 1992). Since 1.4\ GHz measurements at higher 
redshifts sample higher frequencies that are not sensitive to 
these absorption effects, we adopt the standard value $\alpha_{r}=-0.8$
from \markcite{condon92}Condon (1992) to extrapolate
local 8.4\ GHz fluxes to 1.4\ GHz fluxes, in order to
be consistent with high redshift observations.

The ratio of Eqs.~\ref{smmlong} and \ref{radio} with $T_d=47\ $K gives

\begin{equation}
{S_{353\ {\rm GHz}}\over S_{1.4\ {\rm GHz}}} = 1.1\times (1+z)^{3.8}
\label{ratio}
\end{equation}

\noindent
A relation of this type was obtained by CY99
based on the SED of M82. In their relation, the factor of 1.1 in
Eq.~\ref{ratio} is replaced by 0.25 and the power-law index 
$\alpha_s - \alpha_r=3.8$ is replaced
by 4.3. However, M82, with more than 
an order of magnitude lower luminosity than Arp~220
($L_{FIR}=1.78\times 10^{10}\ h_{65}\ {\rm L_\odot}$;
\markcite{hughes97}Hughes, Dunlop, \& Rawlings 1997), 
is not a ULIG, and so we would argue that our result is more 
appropriate for the distant submillimeter sources.
In addition to their analytic result, CY99
give empirical results for the flux ratio versus redshift for
both Arp~220 and M82. Other authors have used the span of the four 
CY99 curves in their Fig.~1 to estimate redshift ranges; 
however, the use of M82 again introduces
an inappropriate uncertainty in obtaining redshifts.

Inverting Eq.~\ref{ratio}, the value of $z$ can be related to the flux 
ratio by

\begin{equation}
z+1=0.98\times \Biggl({S_{353\ {\rm GHz}}\over
S_{1.4\ {\rm GHz}}}\Biggr)^{0.26}
\end{equation}

\noindent
which holds for $z\lesssim3$. For higher redshifts the ratio
of Eqs.~\ref{smmexact} and \ref{radio} must be used.
The flux ratio method for estimating redshifts has the advantage
that the luminosity distance drops out, and thus there is no
dependence on ${\rm H_o}$ or on the cosmology.

We make the assumption that unevolved local ULIGs are representative
of the distant ULIG population.
We show in the following that the Arp~220 SED provides a good
representation of the submillimeter-to-radio flux ratios of
the ensemble of local ULIGs placed at appropriate redshifts.
We use a compilation of 
submillimeter measurements at 3000, 857, 667, and 375\ GHz from
\markcite{rig96}Rigopoulou et al.\ (1996) to infer
353\ GHz fluxes at redshifts of 7.5, 1.44, 0.9, and 0.06.
Fluxes at 1.4\ GHz were extrapolated from the 8.4\ GHz measured values 
of \markcite{condon91}Condon et al.\ (1991)
using a standard synchrotron emission spectrum with
$\alpha=-0.8$, as justified earlier. We plot the mean 353\ GHz
to 1.4\ GHz flux ratios versus redshift as solid diamonds
in Fig.~\ref{figulig}a with ranges from the minimum to maximum flux
ratios. The number of objects in the bins are, in order of decreasing
redshift, 25, 2, 4, and 7.
The data are to be compared with the redshifted Arp~220 flux
ratio calculated from Eqs.~\ref{smmexact} and \ref{radio} and plotted as
a solid line in the figure. The spread in measured ratios
is about a multiplicative factor of two relative to Arp~220,
as indicated by the dashed curves. Thus, the redshifted Arp~220 flux ratio 
is likely to approximate the flux ratios of ULIGs
at high redshifts to this accuracy.

The above conclusion is strengthened by Fig.~\ref{figulig}b where
we have plotted as solid (star forming galaxies) and
open (AGN) squares available submillimeter observations with
corresponding radio and redshift information.
The star forming galaxies fall
neatly on the curves. 
The average submillimeter-to-radio flux ratio of the galaxies
in our survey which have spectroscopic redshifts and have been
observed in the submillimeter are shown in two redshift bins 
as the large filled diamonds with $1\sigma$ uncertainties. 
The lowest point is consistent with a $3\sigma$ null detection, 
but in the higher redshift bin there is a strong
positive detection consistent with the Arp~220 ratio, showing that
these objects have cool dust emission and obey the FIR-radio
relation at these redshifts.
The short horizontal lines show the redshift ranges
for our six significant radio-selected submillimeter detections;
all are in the $z=1-3$ redshift range.
The top portion of Table~\ref{tab3} 
gives the fluxes and the millimetric redshifts and ranges for these six
sources, as well as their bolometric luminosities for both $\Omega_\Lambda=0$
and $\Omega_\Lambda=2/3$ (see \S~\ref{secfluxz}). 

\begin{deluxetable}{crrrrrrr}
\tablewidth{0pt}
\tablenum{3}
\tablecaption{Submillimeter Source Redshifts and Luminosities\label{tab3}}
\tablehead{
\colhead{Catalog} & \colhead{$S_{353\ {\rm GHz}}$} &
\colhead{$S_{1.4\ {\rm GHz}}$} &
\colhead{$z_{min}$} & \colhead{$z$} & \colhead{$z_{max}$} &
\colhead{$L_{bol}$} & \colhead{$L_{bol}$} \\
\colhead{\#} & \colhead{(mJy)} & \colhead{(mJy)} & & & &
\colhead{$(10^{12}\ {\rm L_\odot})$} & \colhead{$(10^{12}\ {\rm L_\odot})$}\\
& & & & & &
\colhead{$\Omega_\Lambda=0$} & \colhead{$\Omega_\Lambda=2/3$}
}
\startdata
33 & 11.9 & 0.124 & 1.0 & 1.5 & 2.1 & 5.0 & 9.1 \\
49 & 10.7 & 0.324 & 1.6 & 2.3 & 3.1 & 4.5 & 9.0 \\
7 & 7.8 & 0.151 & 1.3 & 1.8 & 2.5 & 3.3 & 6.3 \\
11 & 7.5 & 0.148 & 1.2 & 1.8 & 2.4 & 3.2 & 6.0 \\
13 & 7.1 & 0.071 & 1.8 & 2.4 & 3.3 & 3.0 & 6.0 \\
46 (HDF850.2) & 2.4 & 0.046 & 1.3 & 1.8 & 2.5 & 1.0 & 1.9 \\
\\
\nodata & 19.0 & $<0.040$ & $>3.6$ & $>5.0$ & $>7.6$ & $>9.2$ & $>20$ \\
\nodata & 8.8 & $<0.040$ & $>2.5$ & $>3.5$ & $>4.8$ & $>4.2$ & $>8.9$ \\
(HDF850.1) & 5.4 & $<0.040$ & $>2.0$ & $>2.8$ & $>3.8$ & $>2.6$ & $>5.4$ \\
\enddata
\end{deluxetable}

The ULIG sources plotted in Fig.~\ref{figulig}
have a range of emissivity indices, $\beta$, and
dust temperatures in the vicinity of 50\ K. 
It is therefore quite remarkable that all the sources
follow a constrained envelope around the Arp~220 submillimeter-to-radio
flux ratio. The implication of this empirical correlation is that for
practical purposes a temperature-redshift degeneracy 
(\markcite{b99}Blain 1999) does not present a serious problem
to millimetric redshift estimation.

A recent empirical analysis by \markcite{cy00}Carilli \& Yun (2000),
which appeared subsequent to the submission of our paper, finds
conclusions that are consistent with our analysis above.

\section{Radio and Submillimeter Fluxes versus Redshift}
\label{secfluxz}

In our analysis we assume a flat Universe, 
$\Omega_{\rm M}+\Omega_\Lambda=1$,
where $\Omega_{\rm M}$ is the matter density and $\Omega_\Lambda$ the
vacuum density.
The absolute flux values in Eqs.~\ref{smmexact}
and \ref{radio} depend on the cosmology through the luminosity distance
(\markcite{carroll92}Carroll, Press, \& Turner 1992)

\begin{equation}
d_L(z)=c{\rm H_o}^{-1}(1+z)\int_0^z dz'/
[(1+z')^2 (1+\Omega_{\rm M} z') - z'(2+z')\Omega_\Lambda]^{1/2}
\end{equation}

\begin{figure*}[tb]
\figurenum{7}
\centerline{\psfig{figure=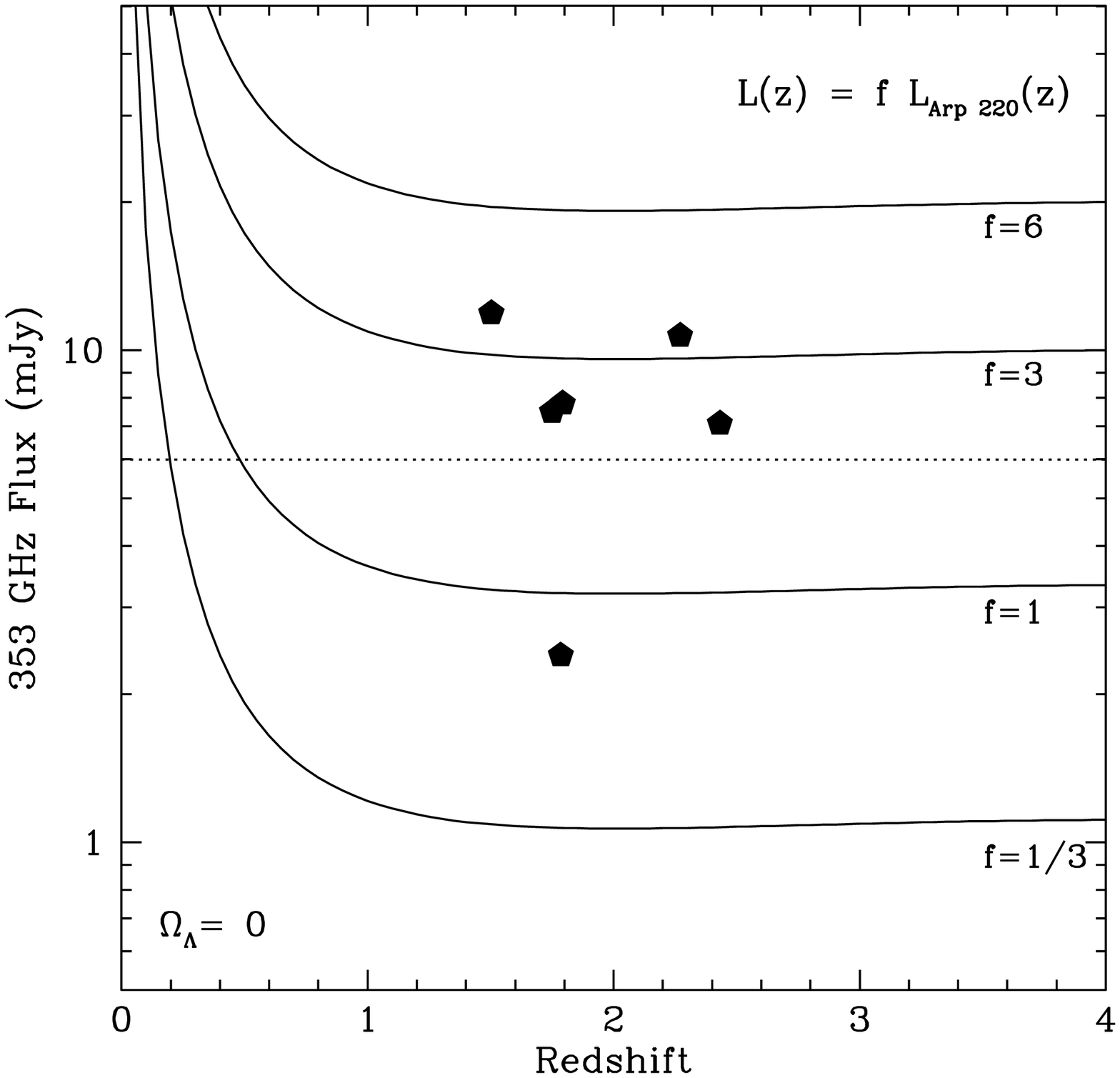,width=3.5in}
\psfig{figure=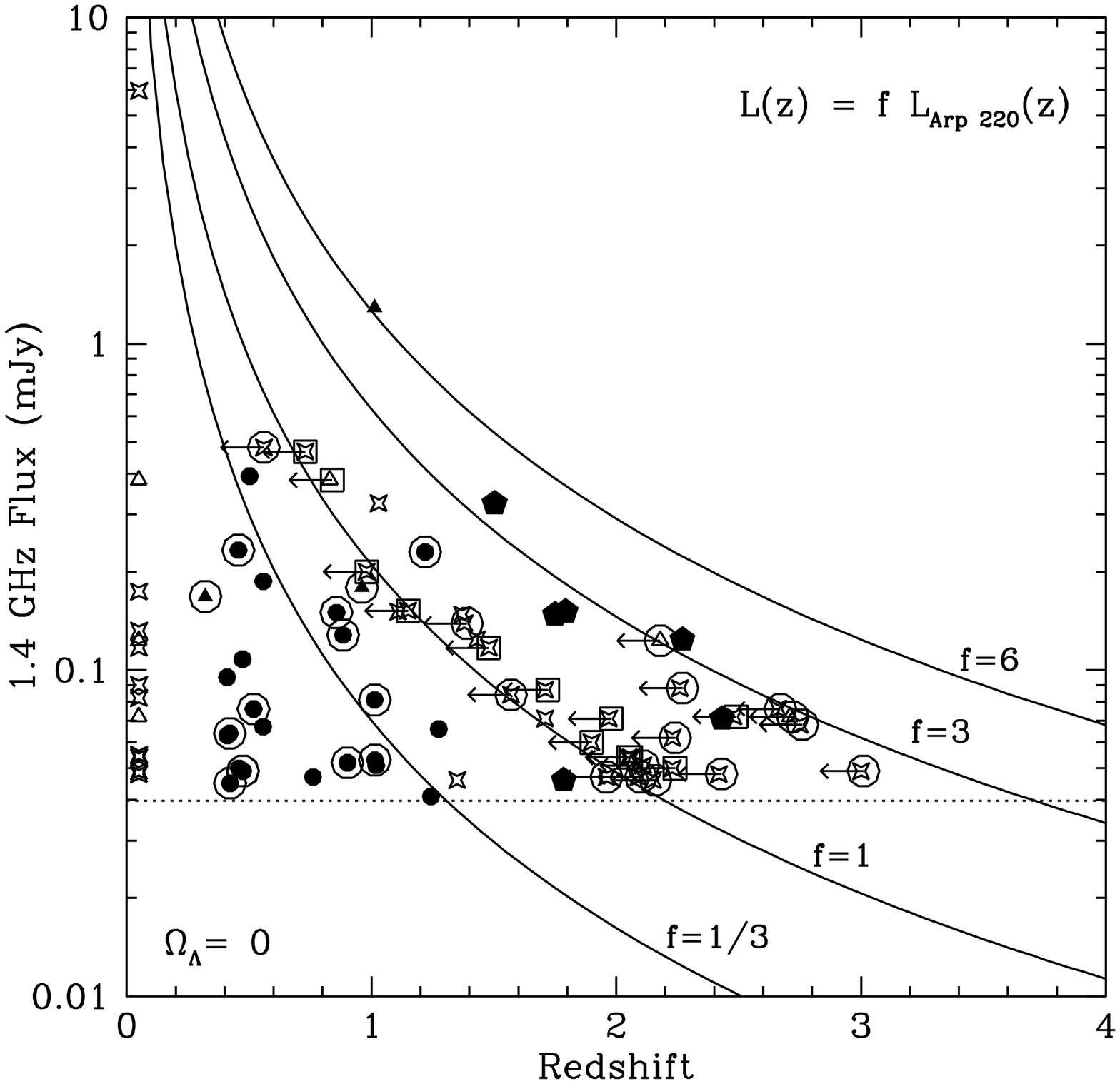,width=3.5in}}
\figcaption[]{
(a)\ Submillimeter flux versus redshift, and
(b)\ radio flux versus redshift for our radio sample.
Solid curves show the predicted flux versus redshift for Arp~220
scaled by $f=1/3$, 1, 3, and 6.
Detection limits of 6\ mJy ($3\sigma$) for the submillimeter and
0.04\ mJy ($5\sigma$) for the radio are indicated with dotted lines.
Filled circles have spectroscopic redshifts while
open crosses do not. Filled (with redshifts) and open triangles have
$\alpha_r>-0.3$.
Solid pentagons are distant ULIGs at $S_{353}>6$\ mJy (our survey)
and $S_{353}<6$\ mJy (H98), shown at their millimetric redshifts.
Open squares (optical/NIR-faint) and open circles (optical/NIR-bright)
were observed but not detected at 353\ GHz at the
6\ mJy ($3\sigma$) level; millimetric redshift limits from the
$1\sigma$ upper limits on the submillimeter measurements
are indicated by the leftward pointing arrows.
Here $\Omega_{\rm M}=1$, $\Omega_\Lambda=0$ with ${\rm H_o}=65$.
\label{figfalloff}
}
\end{figure*}

\begin{figure*}[tb]
\figurenum{8}
\centerline{\psfig{figure=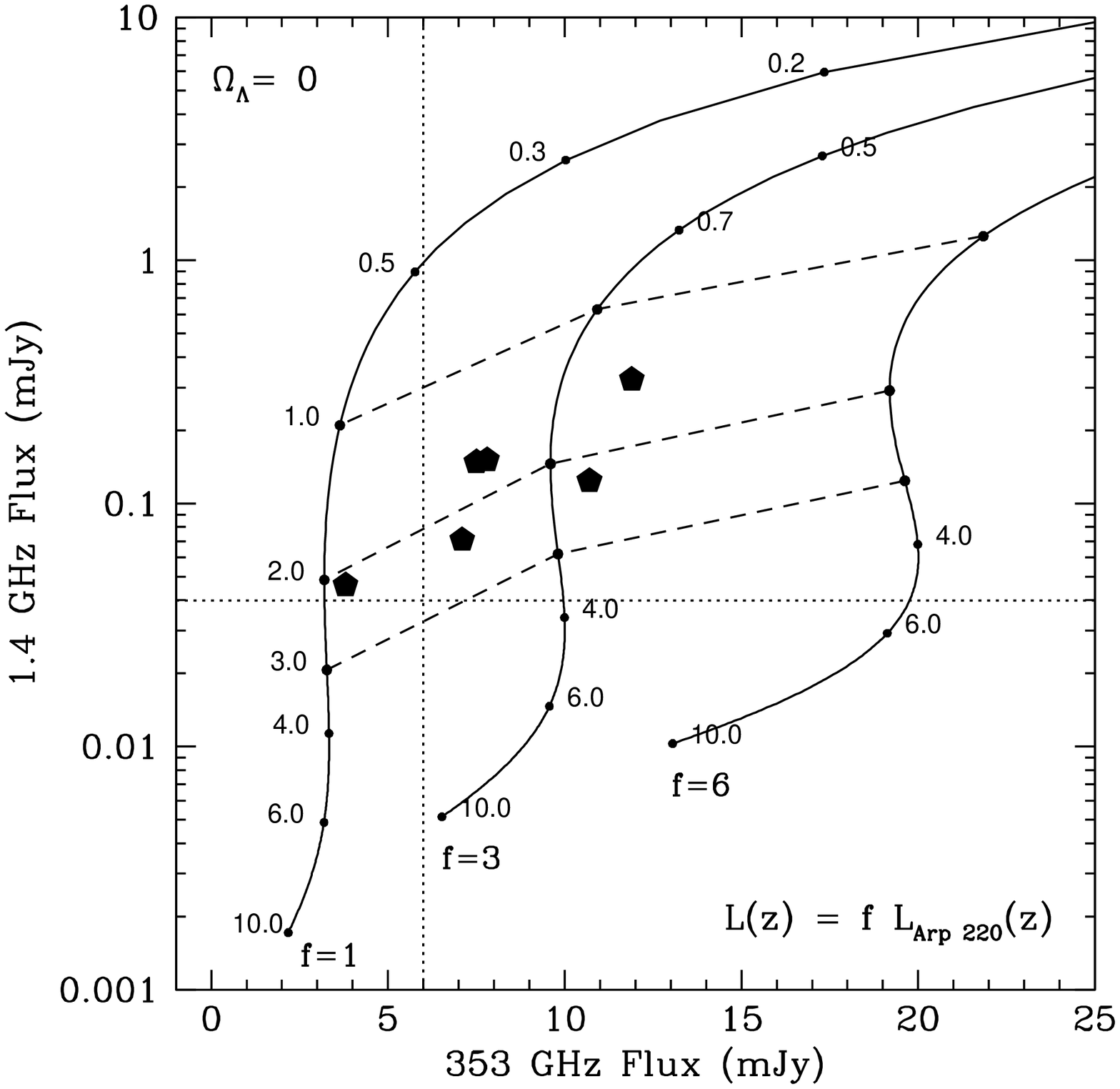,width=3.5in}
\psfig{figure=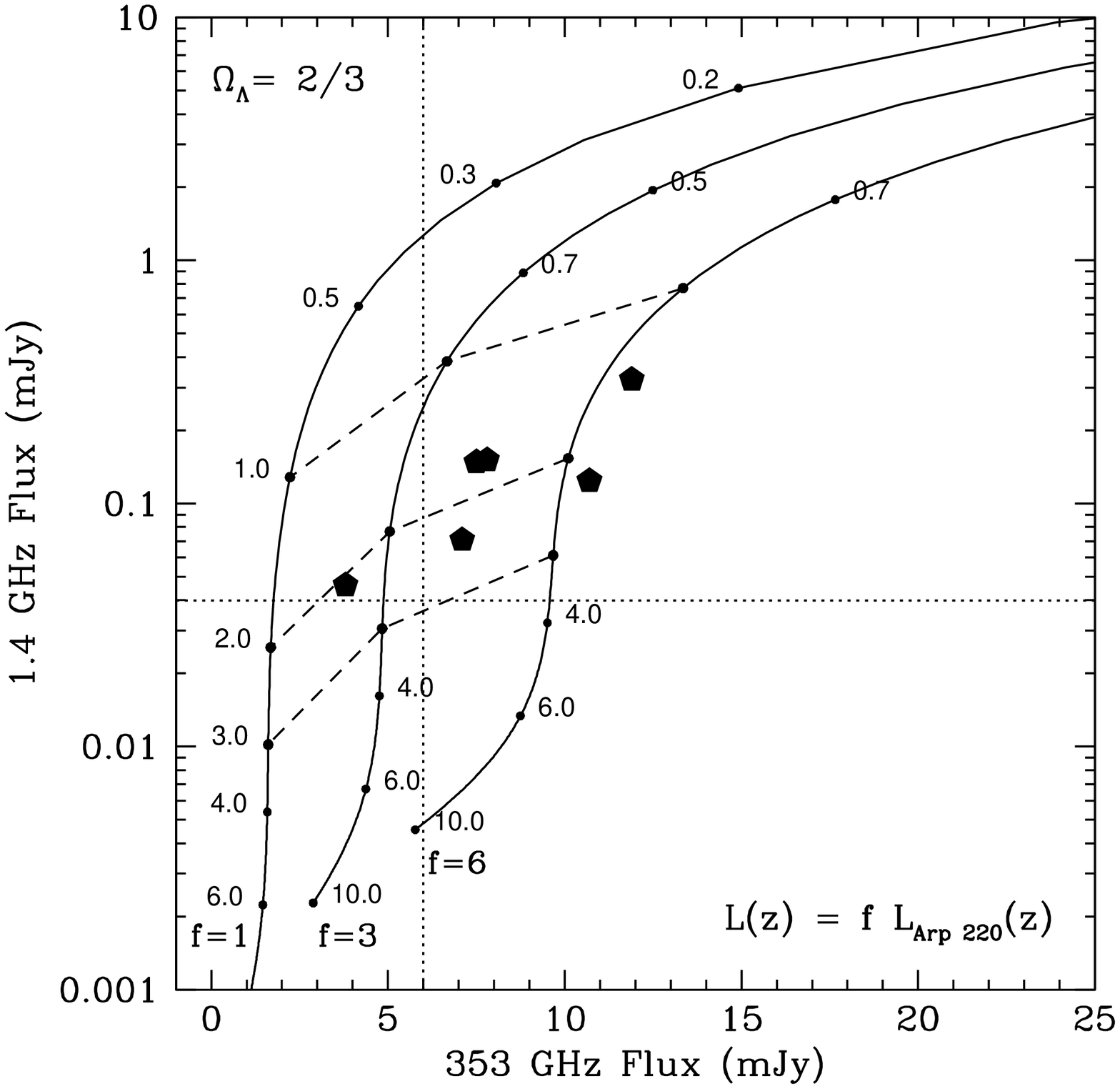,width=3.5in}}
\figcaption[]{
Radio versus submillimeter flux for a redshifted Arp~220
SED scaled by $f=1$, 3, and 6 with (a)\ $\Omega_{\rm M}=1$,
$\Omega_\Lambda=0$, and
(b)\ $\Omega_{\rm M}=1/3$, $\Omega_\Lambda=2/3$.
Redshift labels are marked on the curves.
Dashed lines connect the curves at constant redshifts of $z=1$, 2, and 3.
Dotted horizontal line is the $0.04$\ mJy ($5\sigma$) flux
limit for the 1.4\ GHz sample.
Dotted vertical line is the 6\ mJy ($3\sigma$) limit
for our radio-targeted submillimeter survey.
Symbols are as in Fig.~\ref{figfalloff}.
Approximate redshifts for the sources can be inferred from
the curves and are all in the range $z=1-3$, as demarcated in the
figure by dashed lines.
\label{figsvss}
}
\end{figure*}

\noindent
The basic physics underlying the joint use of radio and submillimeter
observations is contained in Figs.~\ref{figfalloff} and \ref{figsvss}.
Figure~\ref{figfalloff}
shows predicted Arp~220 radio and submillimeter fluxes versus
redshift for the $\Omega_{\rm M}=1$, $\Omega_\Lambda=0$ cosmology
(${\rm H_o}=65$\ km\ s$^{-1}$\ Mpc$^{-1}$) with selected 
overall strengths $f=1/3$, 1, 3, and 6 relative to Arp~220.
$S_{353\ {\rm GHz}}$ is flat with redshift for
$z\gtrsim 1$ due to the negative $K$-correction, whereas
$S_{1.4\ {\rm GHz}}$ falls sharply with increasing redshift due
to the $1/d_L^2$ dependence. 
These behaviors are characteristic of other cosmologies as well.
Thus, radio surveys detect a high
proportion of low redshift sources whereas submillimeter surveys offer
comparable sensitivity to both moderate ($z\sim 1$) and high redshifts.

%
%

%
%

In Fig.~\ref{figsvss}
we illustrate the dependence on cosmology of the absolute fluxes in
the $S_{1.4\ {\rm GHz}}$ versus $S_{353\ {\rm GHz}}$ plane.
Figure~\ref{figsvss}a shows predicted Arp~220 fluxes for the
$\Omega_\Lambda=0$ cosmology, and Fig.~\ref{figsvss}b shows the same
for the $\Omega_\Lambda=2/3$ cosmology
favored by recent distant supernovae type Ia observations
(\markcite{perl99}Perlmutter et al.\ 1999;
\markcite{riess98}Riess et al.\ 1998).
Redshift labels are given on the $f=1$, 3, and 6 curves.
In Fig.~\ref{figsvss}a the bright submillimeter sources fall near the
$f=3$ contour, indicating that these sources have luminosities
several times the luminosity of the Arp~220 prototype.
For the $\Omega_\Lambda=2/3$ cosmology we would infer
somewhat higher luminosities from the curves, although the redshift
estimates remain the same.

From Fig.~\ref{figsvss}, we see that
any submillimeter sources above our detection threshold of 6\ mJy
would not be detectable in the radio if $z\gtrsim 4$.
Of the significant submillimeter sources detected in our survey,
two had no radio counterparts and one, HDF850.1, had only a supplemental
8.5\ GHz radio detection. Based on the millimetric redshift estimator, 
these sources are potentially at very high
redshift. The bottom three lines of Table~\ref{tab3} give the fluxes and 
the millimetric redshifts and ranges, along with
the bolometric luminosities for $\Omega_\Lambda=0$
and $\Omega_\Lambda=2/3$,
for the three significant SCUBA detections without corresponding
1.4\ GHz detections; here the numerical entries are based on the
0.040\ mJy ($5\sigma$) radio flux limit, which corresponds to the 
lowest possible redshift and $L_{bol}$.

Since we have covered only one-third of our $\sim 80$\ square arcminute 
area with our targeted SCUBA observations, there could in principle be 
more high redshift submillimeter sources with no radio detections.
However, our present radio counts already saturate
the bright submillimeter counts distribution from the combined
blank field submillimeter surveys (see Fig.~\ref{figcounts1})
and hence argue against the probability of finding many such
additional sources.

In contrast, sources of Arp~220 or sub-Arp~220 strength
could be seen in the radio but not in the submillimeter with
our present 6\ mJy threshold if $z\lesssim 2$.
Figure~\ref{figfalloff}b shows
that we have many such candidates.
Our targeted SCUBA observations of optical/NIR-faint radio sources
are therefore selecting the high redshift end of the faint radio
source population.

\section{Rest-frame Color versus Redshift}

\begin{figure*}[tbh]
\figurenum{9}
\centerline{\psfig{figure=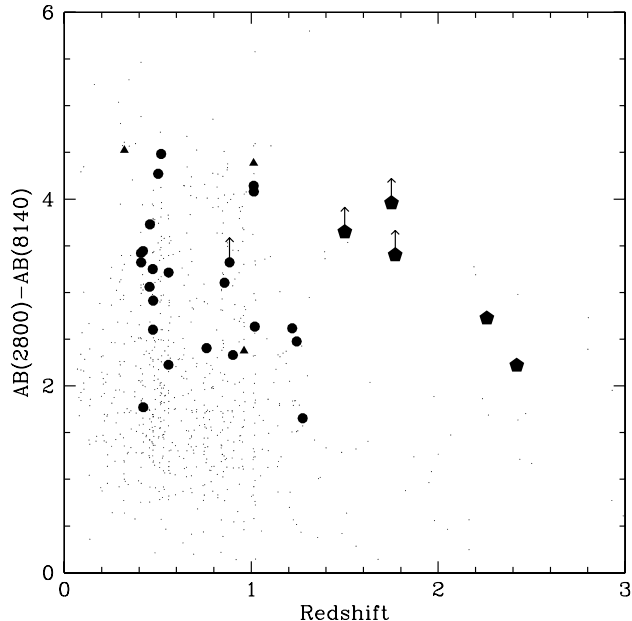,width=3.5in}}
\figcaption[]{
AB(2800)--AB(8140) color versus redshift for the
1.4\ GHz sources with spectroscopic identifications
(solid circles). Solid triangles have $\alpha_r>-0.3$.
Solid pentagons are submillimeter
sources with $HK'$ detections and millimetric redshift estimates.
Tiny circles are from LRIS redshift surveys of the HDF and SSA22
fields and illustrate
how much redder the radio sources are than the typical field galaxy.
\label{figrest}
}
\end{figure*}

In Fig.~\ref{figrest} we plot rest-frame AB(2800)--AB(8140)
color versus redshift for the radio sample with spectroscopic
identifications. The colors of field galaxies in the HDF and SSA22
fields are indicated with tiny solid circles to illustrate how
the radio galaxies (solid triangles for sources with $\alpha_r>-0.3$
but otherwise solid circles) generally have very red AB colors. These
colors range from $\sim 1.5$ to $\sim 4.5$. 
There is no evidence for a color trend with redshift.
The rest-frame AB colors of the submillimeter galaxies
with $HK'$ detections (solid pentagons), although mostly lower limits,
are not inconsistent with the colors of the lower-redshift radio sources.

At $z=2$, AB(2800)--AB(8140) roughly corresponds to AB($I$)--AB($HK'$),
and thus submillimeter sources with similar colors to the radio
sources would have observed Vega-based
$I-HK'$ colors in the range $3.1-6.1$, which would mostly place them 
in the extremely red object (ERO) category 
($I-HK'>3.7$ or $I-K>4$; e.g., 
\markcite{mccracken99}McCracken et al.\ 1999 and references therein).
This result is consistent with
the recent detections of bright ($K<20$) EROs as submillimeter sources
(\markcite{cimatti98}Cimatti et al.\ 1998;
\markcite{dey99}Dey et al.\ 1999;
\markcite{smail99a}Smail et al.\ 1999a).
In general, however, the submillimeter sources are
so optically faint that identifying them as EROs is difficult.

%
%

\section{Properties of the Radio and Submillimeter Populations} 

\begin{figure*}[tbh]
\figurenum{10}
\centerline{\psfig{figure=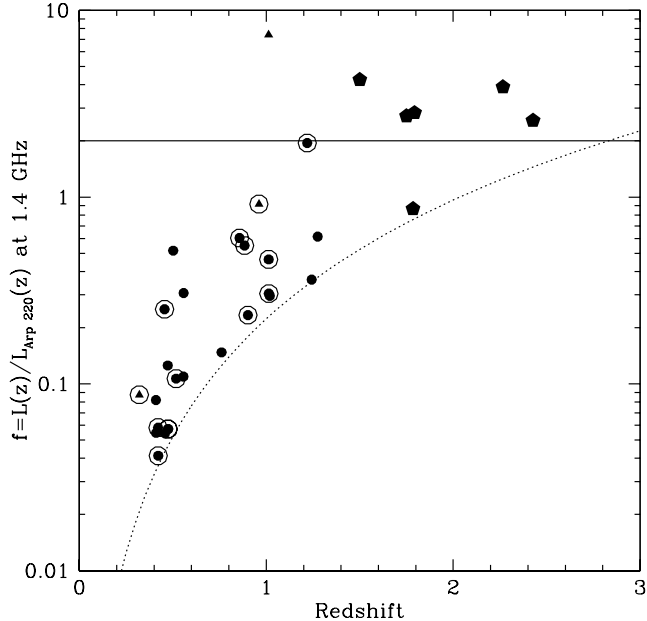,width=3.5in}}
\figcaption[]{
1.4\ GHz source luminosity, relative to the redshifted 1.4\ GHz
luminosity of Arp~220, versus redshift in the $\Omega_\Lambda=0$
cosmology for the sources that have
been spectroscopically observed. The dotted curve represents
the radio threshold of 40\ $\mu$Jy. The solid curve at
a value of two denotes our 6\ mJy submillimeter survey's
effective minimum sensitivity to sources that have two times the
luminosity of Arp~220; the submillimeter source lying below this
value is HDF850.2 of H98. The symbols are as in Fig.~7.
\label{figradiolum}
}
\end{figure*}

The luminosities of our submillimeter sources were obtained by scaling
the Arp~220 luminosity at redshift $z$ by the relative source 
strengths, $f$; these are given in Table~\ref{tab3}.
The FIR luminosity provides a direct measure of the current star
formation rate (SFR);
with present detection capabilities, the radio provides greater
sensitivity to SFRs at low redshifts
($z\lesssim 1.5$) while the submillimeter is superior at high redshifts.
In this sense radio and submillimeter data complement each other 
in getting the SFR over the full range of redshifts.
This relative capability is illustrated in Fig.~\ref{figradiolum}
where the strength factors, $f$, as determined from the radio
luminosities relative to redshifted Arp~220 luminosities, 
are plotted versus redshift in the $\Omega_\Lambda=0$ cosmology.
Here the dotted curve represents the radio threshold of 40\ $\mu$Jy,
and the solid line at a value of two represents our
6\ mJy submillimeter threshold of
roughly two times the luminosity of Arp~220 for the
$\Omega_\Lambda=0$ cosmology (see Fig.~\ref{figsvss}).
These two threshold curves cross at $z\sim 2.3$.
Since the bulk of the radio detections are at $z\lesssim 1.3$ 
and have strength factors $f<1$, their non-detection
in the submillimeter is as expected; our
$z\sim 2$ sources are detected in both the radio and submillimeter.
At redshifts $z\gtrsim 3-4$ the submillimeter may provide the
only access to distant ULIGs.

Figure~\ref{figradiolum} also provides a means to establish the
completeness of the ULIG number distribution versus redshift.
Our distant ULIG sample is complete over
the redshift range $z=1-3$ (except possibly for the 4 out of 22
optically-faint radio sources that were not observed) because these
$z=1-3$ sources are above both radio and submillimeter detection
thresholds. At $z\gtrsim 3$ there is the risk that we may
lose sources that fall below the rising 1.4\ GHz threshold
relative to the Arp~220 luminosity at redshift $z$.

%
%

\subsection{Space Density}

The space density of our five significant 
$S_{850\mu{\rm m}}>6$\ mJy submillimeter 
sources in the redshift range $z=1-3$ is

\begin{mathletters}
\begin{eqnarray}
n & = & 3.5^{+2.4}_{-1.5} \times 10^{-5}, \quad \Omega_\Lambda = 0\\
n & = & 1.1^{+0.7}_{-0.5} \times 10^{-5}, \quad \Omega_\Lambda= 2/3
\end{eqnarray}
\end{mathletters}

\noindent
in units $h^3_{65}\ {\rm Mpc}^{-3}$.
In comparison, the space density of local ULIGs
($z<0.15$; \markcite{kim98}Kim \& Sanders\ 1998) with
bolometric luminosities above 
$10^{12}\ h^{-2}_{65}\ {\rm L_\odot}$ is only
$1.7\times 10^{-7}\ h^3_{65}\ {\rm Mpc}^{-3}$ for $\Omega_\Lambda=0$ and
$1.4\times 10^{-7}\ h^3_{65}\ {\rm Mpc}^{-3}$ for $\Omega_\Lambda=2/3$, 
about a factor of 200 and 80 lower than our high redshift results.
The discrepancy in space density is still larger if we compare with
local ULIGs that are comparably luminous to the $>6$\ mJy population.
Thus, enormous evolution in the ULIG population must take
place between $z\sim 2$ and the present. 

It is interesting to note a possible empirical similarity
between the redshift distribution of quasars and that of submillimeter
sources, which might be expected if quasars are a successor
stage to merger processes forming submillimeter sources 
(Sanders \& Mirabel 1996). The 
relative space density of a complete radio-selected quasar
sample (\markcite{shaver98}Shaver et al.\ 1998), which should be
unaffected by dust obscuration, is approximately described
by a Gaussian as a function of redshift, given by

\begin{equation}
n(z)\propto e^{-(z-2.28)^2/1.72}
\label{gauss}
\end{equation}

\noindent
The peaking of this
distribution at $z\sim 2.3$ and the rapid fall-off at low redshift
seems consistent with our measured submillimeter redshift distribution.
Recent ROSAT X-ray studies have found a very high redshift
tail ($z\sim 4$) to the quasar distribution
that may be above the Gaussian expectation (\markcite{hasinger98}Hasinger 1998).
Since the ROSAT sources were not seen in the optical, it is
plausible that they are dust obscured. Similarly, we have found
a few submillimeter sources that plausibly lie at $z\gtrsim 3$.

\subsection{Conversion from Luminosity to Star Formation Rate}

The observed ultraviolet light can most simply be translated
to a metal production rate where, as first pointed out
in \markcite{cowie88}Cowie (1988), there is an extremely
tight relation since both metals and ultraviolet light
are produced by the same massive stars. However, it is
usual (particularly in the FIR) to translate the metal
production rate to a total
stellar mass production rate, which may be more intuitively
interpreted. Unfortunately, this extrapolation 
requires knowledge of the shape 
of the initial mass function (IMF), which is quite uncertain.

In the following we shall assume a Salpeter IMF
($\psi(M)\propto M^{-2.35}$, $0.1-125\ {\rm M}_\odot$).
The SFR is related to the rest-frame 2800\ \AA\ luminosity 
per unit frequency by

\begin{equation}
\dot{M}=\phi\times L_{2800}
\label{mdot1}
\end{equation}

\noindent
with $L_{2800}$ in ergs\ s$^{-1}$\ Hz$^{-1}$ and $\dot{M}$ in
M$_\odot\ {\rm yr}^{-1}$. The Cowie (1988) estimate of $\phi$
from nucleosynthesis arguments and galaxy SED modelling 
is $2.2\times 10^{-28}$ for the above IMF limits. 
Later \markcite{scl89}Songaila, Cowie, \& Lilly (1989) 
suggested that $\phi$ should be about a factor 1.7 lower based 
on stellar synthesis.
\markcite{madau96}Madau et al.\ (1996) obtained a value of
$1.5\times 10^{-28}$ based on the evolutionary models
of \markcite{bc93}Bruzual \& Charlot (1993). 
The relatively invariant estimates
of $\phi$ are a direct consequence of the
nucleosynthesis arguments underlying the modelling; the
primary uncertainty in $\dot{M}$ lies in the stellar IMF.
We adopt the Madau et al.\ value for consistency with other
recent papers that follow this normalization.
In terms of the solar luminosity, the SFR relation is then

\begin{equation}
\dot{M}=\Phi\times \nu L_{2800}/{\rm L}_\odot
\label{mdot2}
\end{equation}

\noindent
where $\Phi=5.3\times 10^{-10}$.

The conversion of the bolometric luminosity to a total stellar
mass production rate is an even more invariant prediction
of the nucleosynthesis arguments since knowledge of the
exact shape of the ultraviolet SED of the galaxies is not needed. 
The bolometric luminosity in units of the solar luminosity is 

\begin{equation}
L_{bol}=BC\times \nu L_{2800}/{\rm L}_\odot
\label{lbol}
\end{equation}

\noindent
where $BC$ is the bolometric correction.
For a flat $S_\nu$ over the wavelength range 912\ \AA\ to 
22000\ \AA, $BC=2.95$.
\markcite{rr97}Rowan-Robinson et al.\ (1997) computed a bolometric
correction of 3.5 from the Bruzual \& Charlot (1993)
models, which we adopt. Then the SFR is related to $L_{bol}$ by

\begin{equation}
\dot{M}=(\Phi/BC) \times (L_{bol}/{\rm L}_\odot) = 1.5\times 10^{-10}\times L_{bol}/L_\odot
\label{mdot3}
\end{equation}

\noindent
$\dot{M}$ would be increased 
by a factor of 3.3 if we had instead assumed a Miller-Scalo IMF over 
the same mass range. 

Equations~\ref{mdot2} and \ref{mdot3} therefore 
provide a self-consistent
description of the mass production rates seen in the optical
and submillimeter. Light which directly escapes the galaxy
can be mapped with the ultraviolet light, while reprocessed
light, which escapes the galaxy in the submillimeter, can be
calibrated with Eq.~\ref{mdot3}.

We now use the Arp~220 bolometric luminosity
$L_{bol}=1.36\times 10^{12}\ h_{65}^{-2}\ \rm{L}_\odot$
(\markcite{klaas97}Klaas et al.\ 1997) and the 8.4\ GHz radio flux of
0.148\ Jy (\markcite{condon91}Condon et al.\ 1991) 
to calibrate the SFRs for our radio sources.
We find $\dot{M}=200\ h_{65}^{-2}\ {\rm M}_\odot\ {\rm yr}^{-1}$;
thus

\begin{equation}
\dot{M}=L_{8.4\ {\rm GHz}} / 6.4\times 10^{27}
\end{equation}

\noindent
where the normalization factor is in units of ergs\ s$^{-1}$\ Hz$^{-1}$.
From the equations of Condon (1992) translated to the same
assumptions about the IMF, we would obtain a normalization
of approximately $1.8\times 10^{27}$ or about a factor of 
3.5 smaller than our result. Our normalization translated
with the $\nu^{-0.8}$ spectral shape to 1.4\ GHz is

\begin{equation}
\dot{M}=L_{1.4\ {\rm GHz}} / 2.7\times 10^{28}
\label{radiosfr}
\end{equation}

\subsection{Contribution of the Radio and Submillimeter Sources to
the Star Formation Rate Density}

Using Eq.~\ref{radiosfr}, we can convert our radio luminosities 
into SFRDs. We know that 95 per cent of the micro-Jansky radio
sources in the HDF region have been resolved at $0.2''$
resolution with the Multi-Element Radio Linked Interferometer
(MERLIN) at 1.4\ GHz. The median angular size for the radio 
emission is $1''-2''$ (Richards 1999a), which
suggests radio emission on galactic or sub-galactic size scales.
Thus, the radio emission in most of these systems may
originate primarily from star formation, although we cannot exclude 
contributions to the radio flux densities from embedded AGN.

We divide our radio sources with spectroscopic redshifts
into two bins, $0.1< z\le 0.7$ and $0.7< z\le 1.3$, excluding
sources with radio spectral indices $\alpha_r> -0.3$ that 
might be AGN. Of the 38 presumed
star forming objects with $HK'<20$ in our sample, 23
have spectroscopic redshifts. Most of the remaining 15 sources
have not been observed and thus may be expected to follow the
same redshift distribution as the identified sources. We therefore calculate the
differential comoving volume over a survey area 
$(23/38)\times 79.4$\ arcmin$^2$.
The volume is integrated from $z_{min}$, the lower redshift limit of 
the bin, to $z_{max}$, either the radio luminosity limit of the survey 
or the upper redshift limit of the bin, whichever is smaller.
The comoving volume element in Mpc$^3$ is

\begin{equation}
\Delta V = {1\over 3} \times A \times 8.46\times 10^{-8}\ 
\times \Biggl[\Biggl({d_L(z_{max}) \over {1+z_{max}}}\Biggr)^3 - 
\Biggl({d_L(z_{min}) \over {1+z_{min}}}\Biggr)^3\Biggr]
\end{equation}

\noindent
where $A$ is the area in arcmin$^2$ and $d_L$ is in Mpc.

We find that for the $0.1<z\le 0.7$ bin, the SFRD in units 
$h_{65}\ \rm{M_\odot}\ \rm{yr}^{-1}\ {\rm Mpc}^{-3}$ is

\begin{mathletters}
\begin{eqnarray}
\rm{SFRD} & = & 0.033^{+0.012}_{-0.009}, \quad \Omega_\Lambda = 0
\label{eqlowradioa} \\
\rm{SFRD} & = & 0.025^{+0.009}_{-0.007}, \quad \Omega_\Lambda = 2/3
\label{eqlowradiob}
\end{eqnarray}
\end{mathletters}

\noindent
and for the $0.7< z\le 1.3$ bin,

\begin{mathletters}
\begin{eqnarray}
\rm{SFRD} & = & 0.048^{+0.020}_{-0.015}, \quad \Omega_\Lambda = 0
\label{eqhiradioa}\\
\rm{SFRD} & = & 0.032^{+0.014}_{-0.010}, \quad \Omega_\Lambda = 2/3
\label{eqhiradiob}
\end{eqnarray}
\end{mathletters}

\noindent
where the uncertainties are Poissonian based on the number of sources.
These represent lower limits to the SFRD since we have
not attempted to correct for the contributions of faint sources below our
flux limits.

The SFRD from our submillimeter sources can be estimated under the
assumption that star formation dominates AGN contributions.
We can calculate the contribution of the $S_{850\mu{\rm m}}>6$\ mJy
submillimeter sources at $z=1-3$ to the SFRD using Eq.~\ref{mdot3}; we find

\begin{mathletters}
\begin{eqnarray}
{\rm SFRD} & = & 0.023^{+0.016}_{-0.010}, \quad \Omega_\Lambda = 0 
\label{eqsfrdsmma}\\
{\rm SFRD} & = & 0.014^{+0.009}_{-0.006}, \quad \Omega_\Lambda = 2/3
\label{eqsfrdsmmb}
\end{eqnarray}
\end{mathletters}

\noindent
Unlike the number distribution versus redshift, the SFRD needs
large corrections for completeness due to the fact that we are
detecting only relatively bright sources here 
(at about six times the luminosity at which the submillimeter
background is primarily resolved), and the distribution $dN/dS$ 
increases rapidly as $S$ decreases. To estimate the completeness 
correction, we assume that the dependences of $N$ on $S$ and $z$ 
factorize, $d^2N/dSdz=g(S)h(z)$. This is a plausible assumption
in the submillimeter where the fluxes are nearly independent of
redshift; nonetheless, in view of the Smail et al.\ (1999b) study, 
this assumption remains to be confirmed.

We can determine the completeness
correction for the SFRD using the empirical number distribution
at 850\ $\mu$m versus $S$

\begin{equation}
dN/dS=3.0\times 10^4\ {\rm deg}^{-2}\ {\rm mJy}^{-1}/S^{3.2}
\label{sfrd}
\end{equation}

\noindent
that describes the measured submillimeter counts above 2\ mJy
(Barger, Cowie, \& Sanders 1999).
We are making the assumption that the flux to $L_{FIR}$ 
conversion based on Arp~220 applies in the low submillimeter flux 
region; the justification is that even the dominant $\sim 1$\ mJy 
population are near-ULIG sources.
The completeness correction over all submillimeter fluxes
is therefore the measured $850\ \mu$m extragalactic background light (EBL)
divided by the $850\ \mu$m light above 6\ mJy.
The $850\ \mu$m EBL measurement of $3.1\times 10^4$\ mJy\ deg$^{-2}$
from Puget et al.\ (1996) and $4.4\times 10^4$\ mJy\ deg$^{-2}$
from Fixsen et al.\ (1998) imply correction factors in the submillimeter
of 11 and 15, respectively. Thus, the estimated total submillimeter 
contribution to the SFRD in units of
$h_{65}\ \rm{M_\odot}\ \rm{yr}^{-1}\ {\rm Mpc}^{-3}$ is 

\begin{mathletters}
\begin{eqnarray}
{\rm SFRD} & = & 0.25^{+0.17}_{-0.11}, \quad \Omega_\Lambda = 0
\label{eqtruesfrdsmma}\\
{\rm SFRD} & = & 0.15^{+0.10}_{-0.06}, \quad \Omega_\Lambda = 2/3
\label{eqtruesfrdsmmb}
\end{eqnarray}
\end{mathletters}

\noindent
where we have used the factor of 11 completeness correction. 

More speculatively, we can determine the SFRD from higher redshift sources
using our two $>6$\ mJy submillimeter sources without radio counterparts.
In this case we use the volume from $z=3-6$ and the actual area
surveyed in the submillimeter. We find 

\begin{mathletters}
\begin{eqnarray}
\rm{SFRD} & = & 0.029^{+0.038}_{-0.019}, \quad \Omega_\Lambda = 0
\label{eqhisfrdsmma}\\
\rm{SFRD} & = & 0.016^{+0.022}_{-0.011}, \quad \Omega_\Lambda = 2/3
\label{eqhisfrdsmmb}
\end{eqnarray}
\end{mathletters}

\noindent
After including the factor of 11 completeness correction, this
becomes

\begin{mathletters}
\begin{eqnarray}
\rm{SFRD} & = & 0.32^{+0.42}_{-0.20}, \quad \Omega_\Lambda = 0
\label{eqhitruesfrdsmma}\\
\rm{SFRD} & = & 0.18^{+0.24}_{-0.12}, \quad \Omega_\Lambda = 2/3
\label{eqhitruesfrdsmmb}
\end{eqnarray}
\end{mathletters}

H98 used photometric redshift
estimates to infer that four of their five
$S_{850\mu{\rm m}}>2$\ mJy sources were in the redshift
range $z=2-4$. While these source identifications were problematic,
it appears likely from the present work that the redshifts do
lie in this rough redshift range. Using our parameters and a scaled Arp~220 SED,
we find that the SFRD for their sources with $\Omega_\Lambda=0$ is
$0.10^{+0.08}_{-0.05}\ h_{65}\ {\rm M_\odot}\ {\rm yr}^{-1}\ {\rm Mpc}^{-3}$.
If we make a completeness correction to include the contribution
below 2\ mJy, we obtain 
SFRD=$0.28^{+0.22}_{-0.14}\ h_{65}\ {\rm M_\odot}\ {\rm yr}^{-1}\ {\rm Mpc}^{-3}$,
in good agreement with our result in Eq.~\ref{eqtruesfrdsmma}.

The presence of a substantial fraction of AGN-dominated ULIG sources would
reduce the above SFRDs. In a recent near-infrared spectroscopic study of
64 local ULIGs, \markcite{veilleux99}Veilleux, Sanders, \& Kim (1999)
found AGN characteristics in $20-25$ per cent of the sample, which
increased to $35-50$ per cent for the sample with $L_{IR}>10^{12.3}\ {\rm L}_\odot$.
Thus, our $>6$\ mJy contributions to the SFRD may need to be reduced by
a factor
$\sim 1.5-2$. However, the lower AGN fraction in fainter ULIGs seen
locally suggests that AGN contamination may be less of an issue for the 
extrapolated SFRD of the whole submillimeter population.

\subsection{Comparison with the Optical Star Formation Rate 
Density Diagram}

The determination of the SFRD from optical observations
has been a subject of intense investigation. Observations
first indicated a rather rapid rise in the SFRD
from $z=0-1$ followed by a
sharp decline at higher redshifts with the peak SFRD being
$z\sim 1.5$ (Madau et al.\ 1996). A recent modification in
the inferred optical SFRD at low redshifts was made by
\markcite{cowie99}Cowie, Songaila, \& Barger (1999),
whose data indicated a more gradual
rise in the SFRD than had previously been found by
\markcite{lilly96}Lilly et al.\ (1996).

It was realized that dust obscuration
effects could result in factors of 3 to 5
(\markcite{pettini97}Pettini et al.\ 1997;
\markcite{meurer99}Meurer, Heckman, \& Calzetti 1999) increases in the SFRD
at high redshift. With these rather uncertain dust corrections
taken into account, it has been argued that the SFRD flattens at a
constant 
$\rm{SFRD}\approx 0.2\ h_{65}\ \rm{M}_\odot\ \rm{yr}^{-1}\ \rm{Mpc}^{-3}$
in the $\Omega_\Lambda=0$ cosmology for $z=1-5$ (Steidel et al.\ 1999).

%
%

\begin{figure*}[tb]
\figurenum{11}
\centerline{\psfig{figure=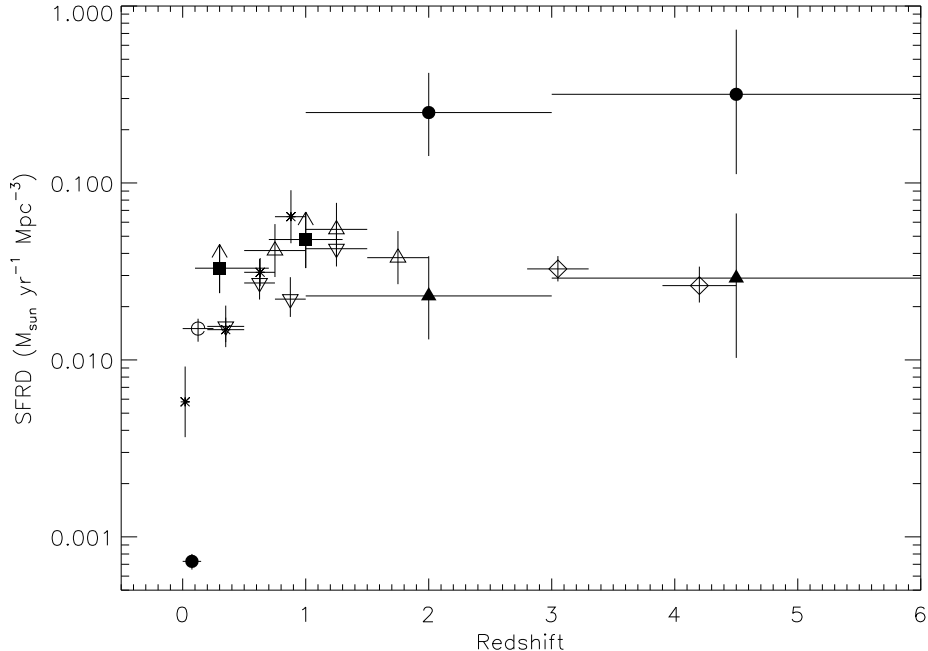,angle=90,width=5.1in}}
\figcaption[]{
SFRD versus redshift for an $\Omega_{\rm M}=1$, $\Omega_\Lambda=0$
cosmology with $H_o=65$. Our submillimeter results for fluxes $>6$\ mJy are
shown as solid triangles, and the results corrected
for completeness are shown as solid circles.
Our radio results are shown as filled
squares. The uncertainties are Poissonian based
on the number of sources. The local point (solid circle) is based
on ULIG data from Kim \& Sanders (1998) and near-ULIG data from
Sanders et al.\ (in preparation).
The open square is from Treyer et al.\ (1998),
asterixes from Lilly et al.\ (1996),
open upside-down triangles from Cowie, Songaila, \& Barger (1999),
open triangles from Connolly et al.\ (1997),
and open diamonds from Steidel et al.\ (1999). All optical
and ultraviolet points are uncorrected for extinction, as
described in the text.
\label{figsfrdvsz}
}
\end{figure*}

In Fig.~\ref{figsfrdvsz} we compare the star formation history 
in the optical (without extinction corrections) 
with that which we obtain in the submillimeter both before
(Eqs.~\ref{eqsfrdsmma} and \ref{eqhisfrdsmma}; solid triangles) 
and after (Eqs.~\ref{eqtruesfrdsmma} and \ref{eqhitruesfrdsmma};
solid circles) correcting for
incompleteness. We also include our newly determined radio
SFRD limits (Eqs.~\ref{eqlowradioa} and \ref{eqhiradioa})
on the figure as solid squares.

The submillimeter contribution to the SFRD inferred from our
$>6$\ mJy observations is comparable to the ultraviolet/optical
contribution to the SFRD.
The two wavelength regimes are likely sampling
different stages in galaxy formation.
The submillimeter detects the formation of massive spheroids
while the ultraviolet/optical detects the formation of smaller disk
or bulge systems.
The approximate equality of the optical and submillimeter
backgrounds supports this hypothesis; the metal density in
present-day disks is roughly comparable to that in the spheroidal
components of galaxies, so comparable amounts of light are expected to
be produced in their formation (\markcite{cowie88}Cowie 1988).

The completeness-corrected submillimeter SFRD, shown by the 
solid circles in Fig.~\ref{figsfrdvsz}, is based on the
assumption that fainter submillimeter sources have the same
redshift distribution and properties as the $>6$\ mJy sample.
Hence the SFRD from the entire population contributing to the
submillimeter background is about an order of magnitude higher
than the extinction-uncorrected ultraviolet/optical SFRD. 
Since the submillimeter measures
reradiated optical light, the observed ultraviolet/optical
SFRD contribution should be added to the submillimeter contribution.

We also plot in Fig.~\ref{figsfrdvsz} 
the SFRD determined from the sum of the SFRD of the 
local ULIG data ($z<0.15$; $L_{bol}>10^{12}\ h^{-2}_{65}\ L_\odot$) 
from Kim \& Sanders (1998) and the local near-ULIG data ($z<0.02$; 
$2\times 10^{11}\ h^{-2}_{65}\ L_\odot<L_{bol}<10^{12}\ 
h^{-2}_{65}\ L_\odot$) 
from Sanders et al.\ (in preparation). The latter $L_{bol}$ selection
is imposed in order to be able to
compare with our completeness-corrected submillimeter data points, 
which should include all sources with luminosities 
$L_{bol}>2\times 10^{11}\ h^{-2}_{65}\ L_\odot$.
We note that the radio data in Fig.~\ref{figsfrdvsz} cannot be 
straightforwardly compared to the ULIG data because the radio sources
may not satisfy the same luminosity criteria.

The completeness-corrected SFRD in
Fig.~\ref{figsfrdvsz} shows a very rapid evolution, 
$\approx (1+z)^{6}$, in the SFRD of ULIGs from $z\sim 0$ to $z\sim 1-3$.
Fast evolution is not surprising if the distant submillimeter
sources are associated with major merger events giving rise to 
the formation of massive spheroidal systems
(Smail et al.\ 1998; Eales et al.\ 1999; Lilly et al.\ 1999; 
\markcite{trentham99}Trentham et al.\ 1999).
Barger, Cowie, \& Sanders (1999) showed that
the volume density of submillimeter sources
at high redshift ($n=5\times 10^{-3}\ h^3_{65}$\ Mpc$^{-3}$ for
$q_o=0.5$) is comparable to the volume density of
present-day elliptical galaxies ($n=10^{-3}\ h^3_{65}$\ Mpc$^{-3}$).

\section{Conclusions}

We have carried out an observational program designed to
establish the overlap between the optical/NIR-faint radio population 
and the submillimeter population and to exploit the complementary
information that the radio and submillimeter wavelengths provide to 
gain insights into the evolution of the galaxy populations.
Our major conclusions are as follows

$\bullet$ We have found that submillimeter sources at bright flux levels 
($> 6$\ mJy) are associated with optical/NIR-faint radio sources 
($40-300\ \mu$Jy). This association
provides a powerful means to conduct submillimeter surveys
by preselecting potential bright submillimeter sources through 
high-resolution deep radio maps. 

$\bullet$ We have shown that the redshifted submillimeter-to-radio
flux ratio of an unevolved Arp~220 SED 
reproduces the ensemble of local ULIG data 
placed at appropriate redshifts, and thus Arp~220 provides a
prototype for ULIG sources at high redshift. From the Arp~220
model we have derived a {\it millimetric} redshift estimator to determine
the redshifts of the submillimeter sources
using the ratio of the submillimeter 
to radio fluxes. Our estimator is consistent with
another recent study by Carilli \& Yun (2000).
The flux strengths relative to the redshifted
Arp~220 SED indicate that the $z=1-3$ bright ($>6$\ mJy) 
submillimeter sources are
$\sim 3\times$ the luminosity of Arp~220 for $q_o=0.5$ and
$\sim 6\times$ the luminosity of Arp~220 for 
$\Lambda$-dominated models. 

$\bullet$ Through radio versus submillimeter flux 
plots of the Arp~220 predictions
with redshift, we have shown that our present
survey is sensitive only to sources with strengths comparable
to or greater than Arp~220 and that our $40\ \mu$Jy radio
threshold precludes the detection of very high redshift sources
($z>4$) if they are radio sources; however, we serendipitously
observed two bright submillimeter sources without radio detections
that may lie at extreme redshifts. An alternate possibility
is that they could be colder galaxies than Arp~220 located at
$z<4$. It remains to be established whether the very faint radio
population down to $1\ \mu$Jy have submillimeter counterparts.

$\bullet$ At $z\lesssim 1$ the ultraviolet/optical contribution to the 
SFRD dominates. At $z>1$ the submillimeter contribution from
$>6$\ mJy sources is comparable to the observed ultraviolet/optical
contribution. The ultraviolet/optical and bright submillimeter are 
likely sampling two different stages in galaxy formation.
The bright submillimeter detects the formation of massive spheroids
while the ultraviolet/optical detects the formation of smaller disk
or bulge systems.

Joint radio/submillimeter surveys are a powerful way to explore
dust-obscured galaxies in the distant Universe since
the radio and submillimeter approaches are complementary and
allow us to infer redshifts and SFRDs for galaxy
sources that we now conclusively see are
inaccessible in shorter wavelength observations.
\vskip 0.5cm

\acknowledgements
We thank Dave Sanders for valuable discussions, Nicolas Biver
for expert observing assistance, and an anonymous referee for helpful
comments about the manuscript.
Support for this work was provided by NASA through Hubble
Fellowship grants HF-01117.01-99A and HF-01123.01-99A awarded by the
Space Telescope Science Institute, which is operated by the
Association of Universities for Research in Astronomy, Inc.,
for NASA under contract NAS 5-26555.

\end{document}